\documentclass{aa}

\usepackage{natbib,twoopt}
\usepackage[breaklinks=true]{hyperref} 
\bibpunct{(}{)}{;}{a}{}{,}             
\makeatletter
  \newcommandtwoopt{\citeads}[3][][]{\href{http://adsabs.harvard.edu/abs/#3}%
    {\def\hyper@linkstart##1##2{}%
     \let\hyper@linkend\@empty\citealp[#1][#2]{#3}}}
  \newcommandtwoopt{\citepads}[3][][]{\href{http://adsabs.harvard.edu/abs/#3}%
    {\def\hyper@linkstart##1##2{}%
     \let\hyper@linkend\@empty\citep[#1][#2]{#3}}}
  \newcommandtwoopt{\citetads}[3][][]{\href{http://adsabs.harvard.edu/abs/#3}%
    {\def\hyper@linkstart##1##2{}%
     \let\hyper@linkend\@empty\citet[#1][#2]{#3}}}
  \newcommandtwoopt{\citeyearads}[3][][]%
    {\href{http://adsabs.harvard.edu/abs/#3}
    {\def\hyper@linkstart##1##2{}%
     \let\hyper@linkend\@empty\citeyear[#1][#2]{#3}}}
\makeatother

\usepackage{array} 
\usepackage{tabularx}
\usepackage{graphicx}
\usepackage[version=3]{mhchem}
\usepackage{txfonts}

\begin{document}

\title{VUV-absorption cross section of carbon dioxide from 150 to 800 K and applications to warm exoplanetary atmospheres}
\author{O. Venot\inst{1, 2}, Y. B\'{e}nilan\inst{1}, N. Fray\inst{1}, M.-C. Gazeau\inst{1}, F. Lef\`{e}vre\inst{3}, Et. Es-sebbar\inst{4}, E. H\'ebrard \inst{5}, M. Schwell\inst{1}, C. Bahrini\inst{1}, F. Montmessin\inst{6},  M. Lef\`{e}vre\inst{1,7},   I. P. Waldmann\inst{8}
}

\institute{Laboratoire Interuniversitaire des Syst\`{e}mes Atmosph\'{e}riques, UMR CNRS 7583, Universit\'{e}s Paris Est Cr\'eteil (UPEC) et Paris Diderot (UPD), Cr\'{e}teil, France\label{LISA}\\
\email{olivia.venot@lisa.u-pec.fr}
\and Instituut voor Sterrenkunde, Katholieke Universiteit Leuven, Celestijnenlaan 200D, 3001 Leuven, Belgium
\and Laboratoire Atmosph\`eres, Milieux, Observations Spatiales (LATMOS), CNRS/IPSL/UPMC, Paris, France
\and Paul Scherrer Institute, Laboratory of Thermal Processes and Combustion, CH-5232 Villigen PSI, Switzerland 
\and School of Physics and Astronomy, University of Exeter, EX4 4QL, Exeter, UK
\and Laboratoire Atmosph\`eres, Milieux, Observations Spatiales (LATMOS), CNRS/IPSL/UVSQ, Guyancourt, France
\and Laboratoire de M\'et\'eorologie Dynamique, UMR CNRS 8539, Institut Pierre-Simon Laplace, CNRS, Sorbonne Universit\'es, UPMC Universit\'e Paris 06, Paris, France
\and University College London, Department of Physics and Astronomy, Gower Street, London WC1E 6BT, UK
}

\date{Received <date> /
Accepted <date>}

\titlerunning{VUV-absorption cross section of CO$_2$}
\authorrunning{Venot et al.}
\abstract{
Most exoplanets detected so far have atmospheric temperatures significantly higher than 300~K. Often close to their star, they receive an intense UV photons flux that triggers important photodissociation processes. The temperature dependency of VUV absorption cross sections are poorly known, leading to an undefined uncertainty in atmospheric models. Similarly, data measured at low temperatures similar to that of the high atmosphere of Mars, Venus, and Titan are often lacking.}
{
Our aim is to quantify the temperature dependency of the VUV absorption cross section of important molecules in planetary atmospheres. We want to provide both 1) high-resolution data at temperatures prevailing in these media and 2) a simple parameterization of the absorption in order to simplify its use in photochemical models. This study focuses on carbon dioxide (CO$_2$).}
{
We performed experimental measurements of CO$_2$ absorption cross section with synchrotron radiation for the wavelength range (115--200 nm). For longer wavelengths (195--230 nm), we used a deuterium lamp and a 1.5~m Jobin-Yvon spectrometer. We used these data in our 1D thermo-photochemical model in order to study their impact on the predicted atmospheric compositions.}
{
The VUV absorption cross section of CO$_2$ increases with the temperature. The absorption we measured at 150~K seems to be close to the absorption of CO$_2$ in the fundamental ground state. The absorption cross section can be separated in two parts: a continuum and a fine structure superimposed on the continuum. The variation of the continuum of absorption can be represented by the sum of three gaussian functions. Using data at high temperature in thermo-photochemical models modifies significantly the abundance and the photodissociation rates of many species, in addition to CO$_2$, such as methane and ammonia. These deviations have an impact on synthetic transmission spectra, leading to variations of up to 5 ppm.}
{
We present a full set of high resolution ($\Delta \lambda$ = 0.03 nm) absorption cross sections of CO$_2$ from 115 to 230 nm for temperatures ranging from 150 to 800 K. A parameterization allows to calculate the continuum of absorption in this wavelength range. Extrapolation at higher temperature has not been validated experimentally and therefore has to be used with caution. Similar studies on other major species are necessary to improve our understanding of planetary atmospheres.}

\maketitle

\section{Introduction}

The more than three thousand exoplanets that have been detected in the last twenty years revealed to us the diversity of worlds that exist in the Universe, in terms of mass, radius, orbital distance, atmospheric composition, etc. Spectroscopic observations performed during transit allow us to characterise the atmospheres of warm short-orbital distance planets whereas direct imaging is used to observe young long-orbital distance planets. Whatever the technique used, the exoplanetary atmospheres that are observable are warm (T$\gtrsim$ 500K). The use of physico-chemical data (e.g. IR molecular line lists, UV absorption cross sections, chemical reaction rates, branching ratios\dots) not corresponding to the high temperatures of these atmospheres yields large sources of error in the understanding of these planets \citep{liang2003source, liang2004insignificance}. In a collaborative white paper, \cite{Fortney2016} point out several areas where experimental work on molecular data at high temperatures is required in order to improve models used in exoplanet science. With the future space- or ground-based telescopes that will be developed in the coming years (JWST, E-ELT\dots), investigating  these research fields becomes urgent.\\
The need for appropriate data to study planetary atmospheres can also be found much closer to us, in our Solar System. In Mars, Venus, or Titan atmospheres for instance, UV radiation penetrates in regions around 150 K. While these bodies have been known and studied for several decades, lots of physico-chemical data at low temperature are still lacking.

We aim at improving models of planetary atmospheres by determining VUV absorption cross sections of important molecules on a large temperature range. In this paper, we focus specifically on carbon dioxide (CO$_2$).
The atmospheres of Mars and Venus being mainly composed (by more than 95 \%) of CO$_2$, this molecule plays a primary role in the photochemistry of these two planets. The VUV absorption cross section of CO$_2$ is therefore essential data to calculate photodissociation rates in photochemical models \citep{lefevre2004, montmessin2011}, and also necessary to analyse the spectrometric data acquired by current spatial instruments such as SPICAM (mission Mars Express), IUVS (mission MAVEN), or SPICAV (mission Venus Express).
Even if exoplanets composed mainly of CO$_2$ have not been detected yet, but must probably exist, the presence of carbon dioxide has been inferred in several observations \citep[e.g.][]{Swain2009HD189, Swain2009HD209, Madhu2009ApJ, Madhu2011Nature}, proving that this species is important in exoplanet atmospheres.

It has already been shown, for a long time, that the VUV absorption cross section depends on temperature. The first experiments performed at temperatures different from 298~K were performed by \cite{Lewis1983297}. They determined the cross section of CO$_2$, $\sigma_{CO_2}(\lambda,T)$, at 200, 300 and 370~K in the wavelength range 120--197 nm. They observed an increase of the absorption cross section with the temperature at long wavelengths. Several years later, \cite{yoshino96b} confirmed this result by measuring $\sigma_{CO_2}(\lambda,T)$ in the range 118.7--175.5 nm at 195 and 295~K. These measurements have been completed by \cite{parkinson2003} and \cite{stark2007} who studied at these two temperatures the ranges 163--200 nm and 106.1--118.7 nm, respectively.\\
Experimental measurements down to 195~K exist but this temperature does not correspond to the $\sim$150~K at which a large fraction of the CO$_2$ photodissociations occurs in Mars and Venus \cite[e.g.][]{Smith2004, zasova2007}. In the absence of data corresponding to this temperature, atmospheric modellers usually adopt one of the following two solutions : 1) Extrapolating the coldest VUV absorption cross section known (i.e. 195~K) to lower temperatures assuming the same variation than between 195 and 298~K. 2) Assuming that no thermal dependance exists for temperature less than 195~K. By simplicity, the second option is the mostly commonly chosen. Subsequently, both methods give a highly uncertain absorption cross section at 150~K. A third method consists of modelling the absorption cross sections using theoretical calculations based on ab initio potentials. To our knowledge it is rarely used by atmospheric modellers.

Concerning the high temperature domain, some measurements have also been performed in the past. \cite{zuev1990uv} measured the absorption of carbon dioxide from 190 to 350~nm between 1000 and 4300~K. Later, \cite{jensen1997ultraviolet} performed measurements in the range (230--355 nm) at four temperatures between 1523 and 2273~K.
\cite{Schulz2002} also acquired data for carbon dioxide at nine temperatures between 880 and 3050~K on the wavelength range (190--320 nm). \cite{oehlschlaeger2004ultraviolet} reached higher temperatures (4500~K) but determined $\sigma_{CO_2}(\lambda, T)$ at four wavelengths only (216.5, 244, 266, and 306~nm). They determined a semi-empirical formula to fit their data. More recently, \cite{venot2013high} published the first VUV absorption data of CO$_2$ for wavelengths lower than 190 nm at high temperatures. They presented measurements performed between 300 and 550~K in the (115--200 nm) wavelength range, as well as data between 465 and 800~K in the 190--230 nm region. From this dataset, an empirical law was determined to calculate the absorption cross section of carbon dioxide between 170 and 230~nm. The comparison of those data with the results published by \cite{Schulz2002} showed a disagreement, which might be due to an overestimation of the temperature in the measurements or an underestimation of the absorbance (see \citealt{venot2013high} for a more detailed discussion). Recently, \cite{Grebenshchikov2016} analyzed the temperature dependence of the absorption cross section of carbon dioxide for wavelengths lower than 250 nm using a first principles model and found a very good agreement with the results of \cite{venot2013high}.\\
We present here a complete dataset for the absorption of CO$_2$ from 150 to 800 K in the range 115--230 nm. We completed the previous measurements of $\sigma_{CO_2}(\lambda, T)$ and improved the data processing. The results presented in \cite{venot2013high} have thus been reevaluated. We fitted all the data with gaussian functions, which allowed us to determine a parameterization of the $\sigma_{CO_2}(\lambda, T)$ continuum valid on the entire wavelength range studied.
In Sect.~\ref{sec:procedures}, we present our experimental setup and our procedures. The results of the measurements are shown in Sect~\ref{sec:results}, and applications to high temperature atmospheric models are presented in Sect.~\ref{sec:application}. Finally, the main conclusions are summarised in Sect.~\ref{sec:concl}.

\section{Experimental Methods and Procedures}\label{sec:procedures}
\subsection{Measurements}

Data have been acquired during three campaigns of measurements : two at the synchrotron radiation facility BESSY\footnote{Berliner Elektronenspeicherring-Gesellschaft f\"ur Synchrotronstrahlung} in July 2011 and June 2014, and one in the Laboratoire Interuniversitaire des Syst\`emes Atmosph\'eriques (LISA) in May 2012. Experimental conditions in terms of temperature, pressure and wavelength studied during each campaign are summarised in Table~\ref{tab:recap}. All the high temperature measurements have been performed on the same experimental setup (described in detail in \citealt{venot2013high}), except that the optical path length varied: 133 cm in 2011, 147 cm in 2012, and 143.9 cm in 2014. To summarise, our setup is composed of a cylinder cell with a length of 120 cm, placed in an oven. At the entrance of the cell, a cross is mounted in order to connect the gas injection disposal and the MKS Baratron\textregistered~ capacitance manometers (range 10$^{-4}$ --1 mbar and 1--1000 mbar). A Mg-F$_2$ window is placed at the beginning of the cross. The other extremity of the cell is closed by a Mg-F$_2$ window and a solar blind photomultiplier measured the VUV radiation intensity. The optical path length is the distance between the two Mg-F$_2$ windows, which represents the absorption cell. The uncertainty on the temperature for the high-temperature measurements is 5\% (see discussion in Sect.~\ref{sec:temp}).
For the measurements at temperatures lower than 300~K, the experimental setup is different and is described in detail in \cite{Ferradaz2009}. Briefly, the cell is surrounded by a double wall in which liquid nitrogen is flowing, which allows the temperature of the gas contained in the cell to decrease. The temperature is measured thanks to a thermocouple fixed on the inside wall of the cell. We observed a temperature gradient in the cell, leading to an uncertainty on the temperature, $\Delta$T~=~$\pm$~5~K. The length of the cell is 108.4 cm, so the total optical path length is 116.8 cm. Uncertainty on the optical path length, $\Delta$l, is lower than 0.1 cm.

\begin{table}[!h]
\caption{Experimental conditions for each campaigns of measurement.} \label{tab:recap}
\centering
\begin{tabular}{ccccc}
\hline \hline
T (K)    & $\lambda$ (nm) & Pressure (mbar) &Period   & Facility  \\
\hline
150 & 115--200 & 0.2--3 & June 2014 & BESSY \\
\hline
170 & 115--200 & 0.3--15 & June 2014 & BESSY  \\
\hline
195 & 115--200 & 0.4--400 & June 2014 & BESSY  \\
\hline
230 & 115--200 & 0.3--491 & June 2014 & BESSY  \\
\hline
300 & 115--200 & 0.25--6.1 & July 2011 & BESSY  \\
\hline
420 & 115--200 & 0.40--7.6 & July 2011 & BESSY  \\
\hline
500 & 115--200 & 0.40--8.2 & July 2011 & BESSY  \\
\hline
585 & 115--200 & 0.53--5.2 & June 2014 & BESSY  \\
  & 195--230 & 73--700 & May 2012 & LISA  \\
  \hline
700 & 115--200 & 0.52--7.4 & June 2014 & BESSY  \\
  & 195--230 & 95--330 & May 2012 & LISA  \\
  \hline
800 & 115--200 & 0.40--8.5 & June 2014 & BESSY  \\
  & 195--230 & 30--398 & May 2012 & LISA  \\
 \hline
\end{tabular}
\end{table}

For all the temperatures studied, spectra were recorded at least at three different pressures between 0.2 and 700 mbar to check the reproducibility of our measurements. Pressure was measured with two MKS Baratron\textregistered ~capacitance manometers, leading to an uncertainty of 1\% corresponding to the precision of the acquisition. The pressure used for each measurement was adapted to the absorption of the gas which varies by several orders of magnitude over the entire wavelength range. This variation forced us to divide the wavelength range into 2 or 3 sub-ranges, in order to obtain acquisitions not saturated, with transmission between 10 and 90\%. The resolution of the data is $\Delta \lambda$ = 0.03 nm.

The total uncertainty on the absorption cross sections is dominated by the uncertainties on the temperature and the pressure of the gas. Comparing the spectra we obtained at different pressure and using a different temperature in the data processing, we estimated it to be about 10\%.

\subsection{Data processing}
\subsubsection{Temperature and wavelength calibration}\label{sec:temp}

In \cite{venot2013high}, we explained that the temperature of the gas is not constant over the entire optical pathway during our high-temperature measurements. This important gradient of temperature can be modelled by a symmetrized inverse exponential function (Eq. 2 in \citealt{venot2013high}). In \cite{venot2013high}, the data were processed considering that the measured absorption corresponded to a gas at a mean temperature $T_{mean}$. This procedure led to a doubtful behaviour of the absorption cross section of CO$_2$ between 134 and 150 nm. In this range only, the high-temperature absorption cross sections were lower than at 300~K. By studying another molecule (NH$_3$, Venot et al. in prep.), we realised that this approximation was not correct. In fact, the absorption is dominated by the portion of the gas at the highest temperature. Thus, we have reprocessed here all our data considering that the gas was at the maximum temperature $T_{max}$ on all the optical pathway. $T_{max}$ depends only on the temperature set by the oven ($T_{set}$) and can be calculated with the formula established in \cite{venot2013high}:
\begin{equation}
T_{max}=0.53\times T_{set} + 181
\end{equation}
With this new procedure, one cannot observe anymore a decrease of the absorption in the range 134--150 nm. All the data presented here have been calculated considering $T_{max}$.
Note that this is applicable only for the measurements at T > 300 K.

Concerning the wavelength calibration, we used the absorption cross sections of CO$_2$ at ambient temperature published by \cite{yoshino96b}, \cite{parkinson2003}, and \cite{stark2007}.

\subsubsection{Carbon monoxide}

We observed the signature of carbon monoxide on some of our highest-temperature raw data (T$_{max}$ > 585 K). The presence of this molecule in our spectra is due to the thermal decomposition of CO$_2$, which is catalysed by wall reactions. Using two CO$_2$ spectra at the same temperature in which CO appeared with different intensities, we isolated the spectral signature of CO. Then, we subtracted the CO feature, applying a correction factor, depending on the temperature and the initial amount of carbon dioxide in the cell. This factor was estimated visually in order to remove completely the CO features, without modifying the normal shape of CO$_2$ spectra. The amount of CO in the cell represents less than 1\% of the total amount of gas.

\subsubsection{Calculation of the absorption cross section}

Absolute photoabsorption cross sections can be calculated using the Beer-Lambert law
\begin{equation}\label{eq:sigmaTconst}
\sigma = \left(\frac{1}{nL}\right) \times \ln \left(\frac{I_0}{I}\right),
\end{equation}
where $\sigma$ is the absorption cross section (cm$^2$), $I_0$ the light intensity transmitted with an empty cell, $I$ the light intensity transmitted through the gas sample, $L$ the absorption path length (cm), and $n$ the volume density of the gas (cm$^{-3}$), following the relation $P=nk_BT$, where $T$ and $P$ are the temperature (K) and the pressure (Pa) of the sample, respectively, and $k_B$ the Boltzmann constant.

\section{Results and Discussion}\label{sec:results}

\subsection{Photoabsorption cross section from 115 nm to 230 nm}
\begin{figure*}[!htb]
\centering
\includegraphics[angle=0,width=1.8\columnwidth]{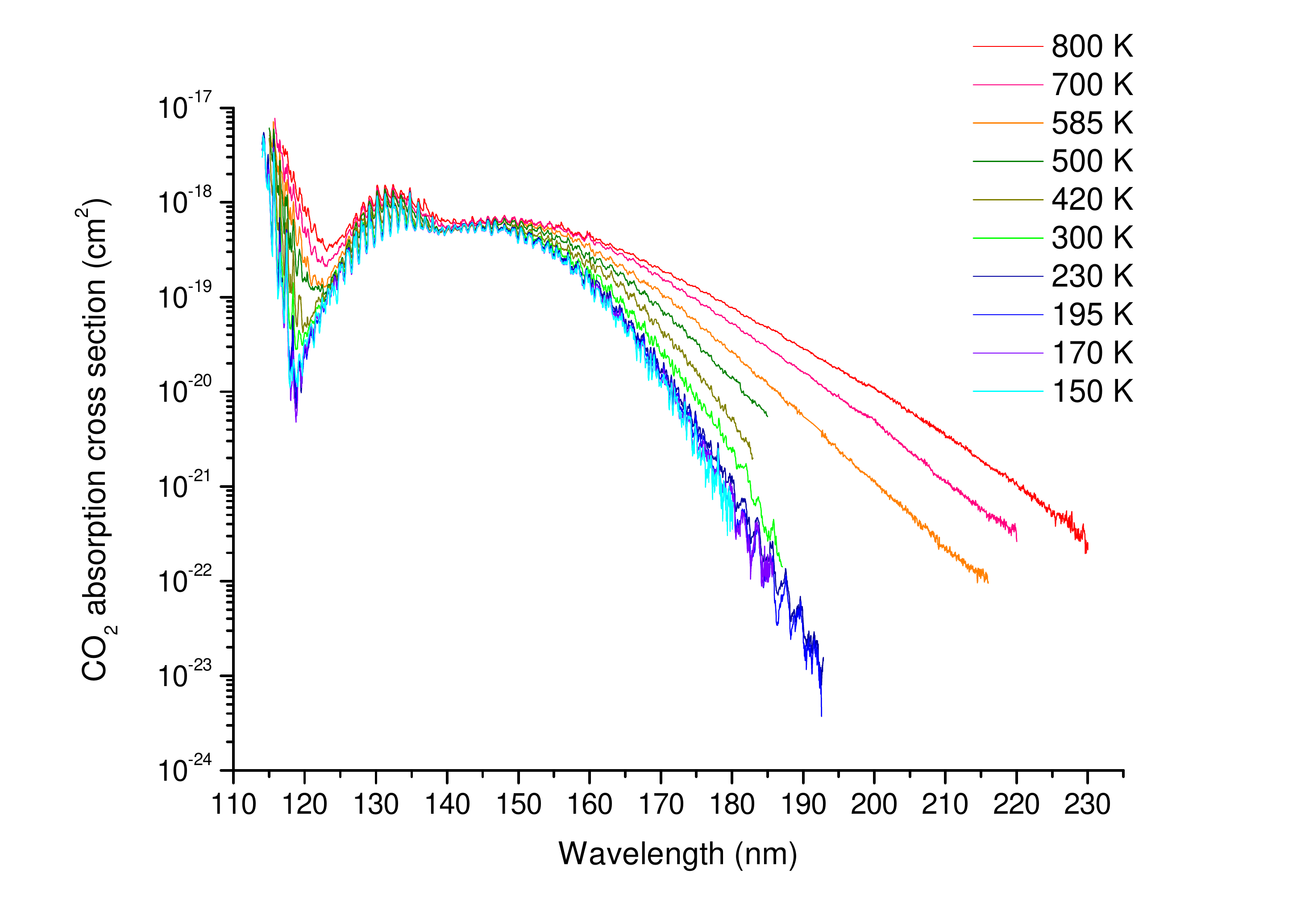}
\caption{Absorption cross section of CO$_2$ (cm$^2$) at temperature between 150 and 800 K.}
\label{fig:crosssection}
\end{figure*}
 
We obtained the absorption cross section of carbon dioxide between 150 and 800 K (Fig.~\ref{fig:crosssection}).
One can notice that the temperature dependency of the absorption depends on the wavelength. Around Lyman $\alpha$, the cross section varies by two orders of magnitude between the two extreme temperatures, whereas between 130 and 150 nm, the variation of the absorption is less important ($<$ factor 10 and sometimes almost identical). For wavelengths $>$160 nm, the deviation of the absorption cross section becomes more pronounced as the wavelength increases.
It can also be noticed that the absorption cross section does not decrease in the same proportion at low temperature than it increases at high temperature. The difference between the four absorption cross sections at low temperature (T$<$300~K) is relatively small. At 175 nm for instance, there is only a factor 3.75 between the ones at 150~K and 300~K. As the temperature decreases, the absorption cross section varies less and less and it seems that a lower limit for the absorption cross section is reached at 150 K. This can be explained by the fact that around this temperature, almost all the molecules are in the vibrational ground state and thus the population of the vibrational levels is not changing anymore. Assuming a Maxwell-Boltzmann distribution between two levels of energy only (i.e. neglecting the higher levels of excitation), one can calculate that 4\% of the population of the fundamental level is in the first energy level ($\nu$=667 cm$^{-1}$) at 300 K, whereas at 150 K, this ratio is lower than 0.2\%.

\begin{figure}[!htb]
\centering
\includegraphics[angle=0,width=\columnwidth]{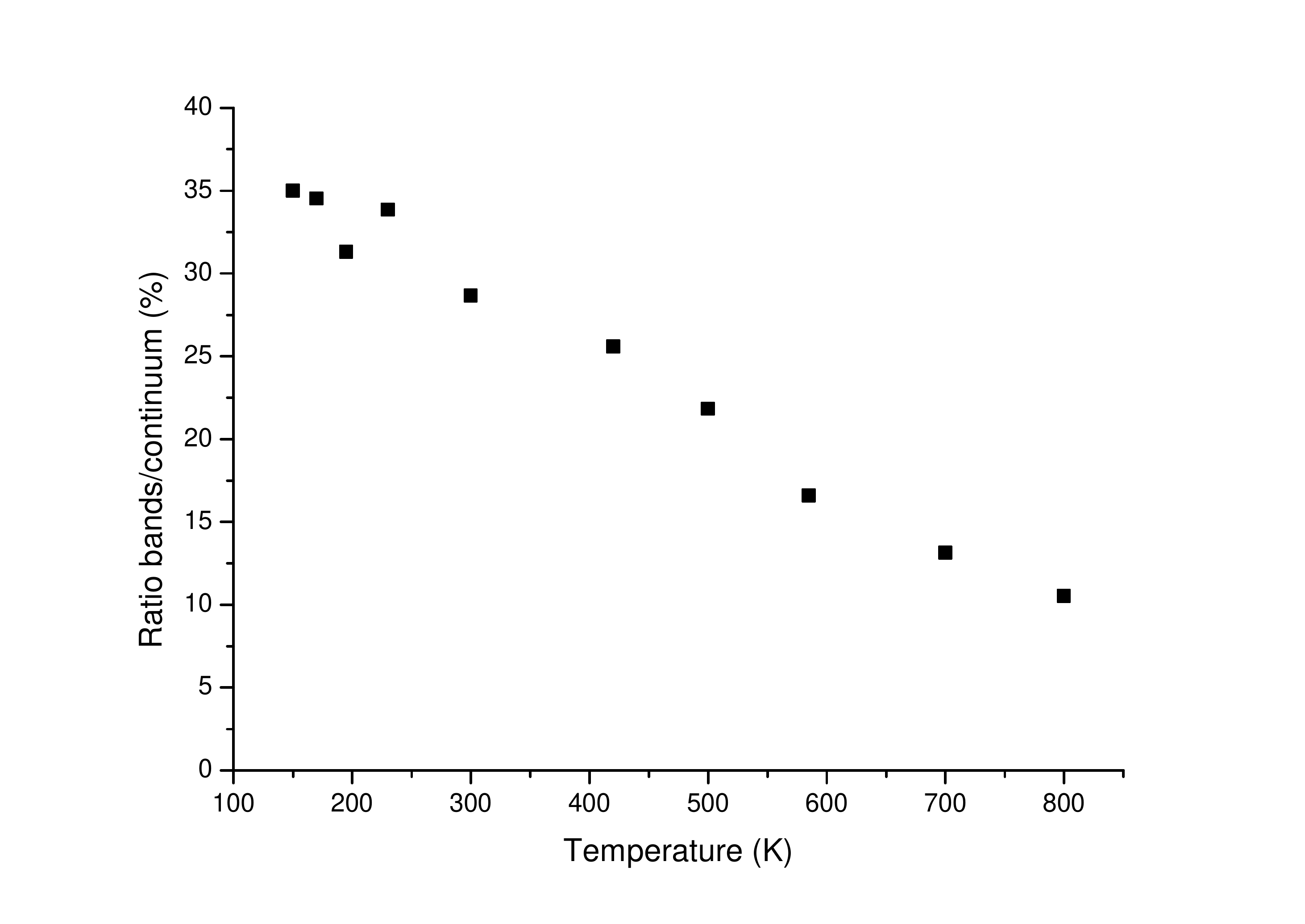}
\caption{Thermal evolution of the ratio between the integrated absorption due to the fine structure and due to the continuum.}
\label{fig:ratio_contrib}
\end{figure}

To analyse in detail the behaviour of the absorption cross section with the temperature, we separated each data set into two parts: the continuum and the fine structure, superimposed on the continuum. The fine structure contains all the small peaks (with a spectral width of $\sim$1nm) corresponding to vibrational energy transitions, and the continuum is the baseline that passes through the minima of all these small peaks.  At each temperature, we integrated the two components over the entire wavelength range and compared these values. As can be seen in Fig.~\ref{fig:ratio_contrib}, when the temperature increases, the contribution of the fine structure to the total absorption decreases. We consider that this contribution of the fine structure becomes negligible for T~>~500 K, as it represents less than 20~\% of the absorption due to the continuum. Note that what we call the "continuum" is in reality overlapping rovibrational transitions, separated by very small energy gaps \citep{grebenshchikov2013photodissociation}.

In this paper, we focus on the analysis of the continuum only, and on the implication of these new data for high temperature atmospheric studies. The analysis of the fine structure is under progress and will be presented in a forthcoming paper, with applications to low temperatures atmospheric studies.

\subsection{Analysis of the continuum of the absorption}

To facilitate the use of these data in atmospheric models, we determined a parameterization to model the absorption cross section continuum. We found that the continuum of the absorption cross section, hereafter $\sigma_{cont}(\lambda, T)$, can be represented as the sum of three gaussian functions :
\begin{equation}\label{eq:sigmatot}
\sigma_{cont}(\lambda, T) = \sigma_1(\lambda,T) + \sigma_2(\lambda,T) + \sigma_3(\lambda,T).
\end{equation}

Our approach is consistent with the fast dissociation of CO$_2$ after the VUV excitation and with the so-called reflection principle (\citealt[p. 392]{herzberg1989molecular}, \citealt[p.110-112]{schinke1995photodissociation}), which essentially approximates the absorption spectrum of each participating electronic state in terms of a Gaussian the parameters of which are controlled by the electronic ground state and the properties of each potential in the Franck-Condon region.

For convenience, we expressed each gaussian function $\sigma_i$ using the wavenumber $\nu$ (cm$^{-1}$), instead of the wavelength $\lambda$ ($\lambda = 1 / \nu$). We obtained the following expression: 
\begin{equation}\label{eq:sigmai}
\sigma_i(\lambda, T) = \sigma_i(\nu,T)=A_i(T) \times \exp\left(-\frac{(\nu-\nu_{ci})^2}{2\times s_i(T)^2}\right),
\end{equation}
where $A_i(T)$ the amplitude (cm$^2$), $s_i(T)$ the spectral width (FWHM) (cm$^{-1}$), and $\nu_{ci}$ the position centre of the gaussian $i$ (cm$^{-1}$).
\begin{figure}[!htb]
\centering
\includegraphics[angle=0,width=\columnwidth]{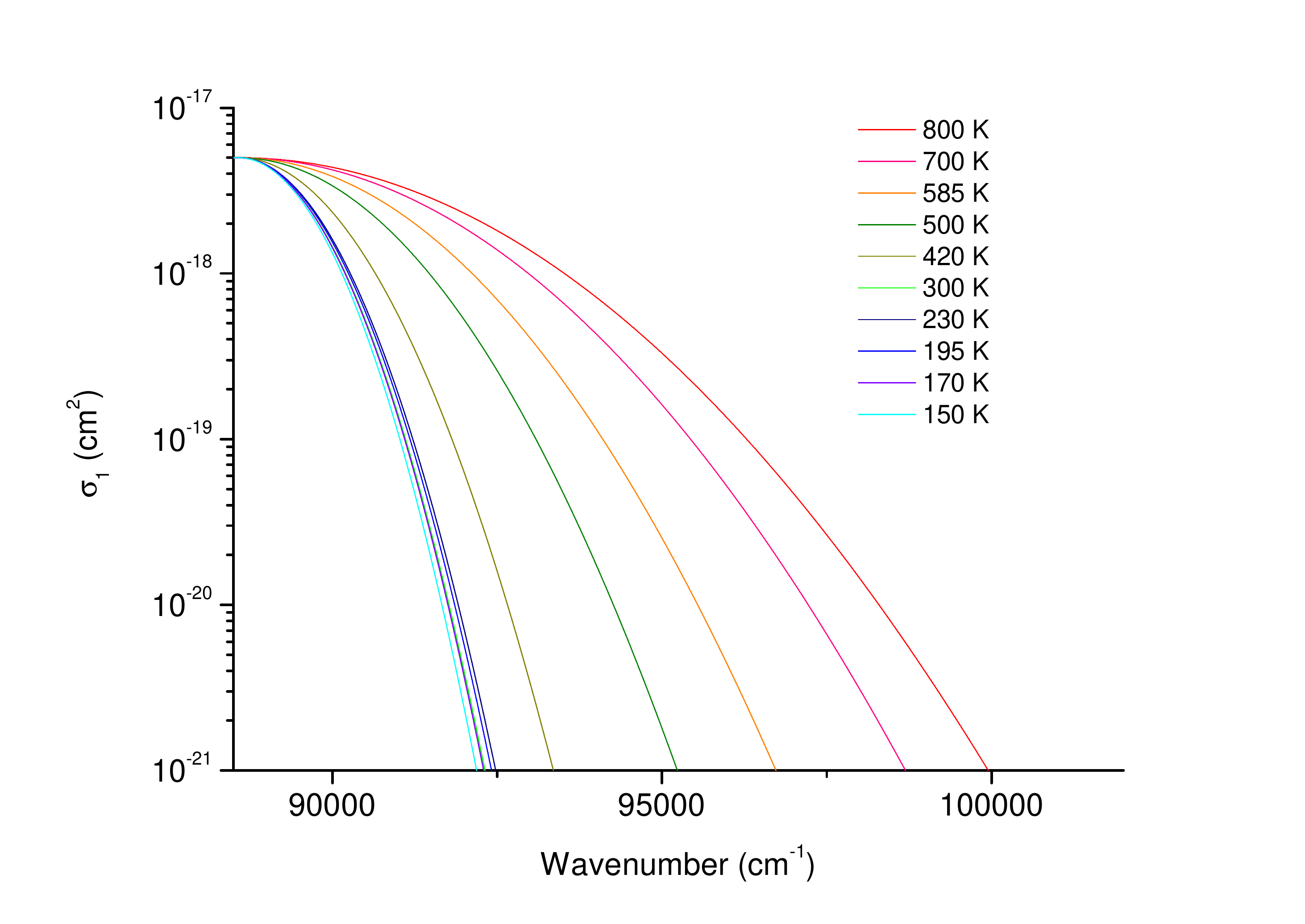}
\includegraphics[angle=0,width=\columnwidth]{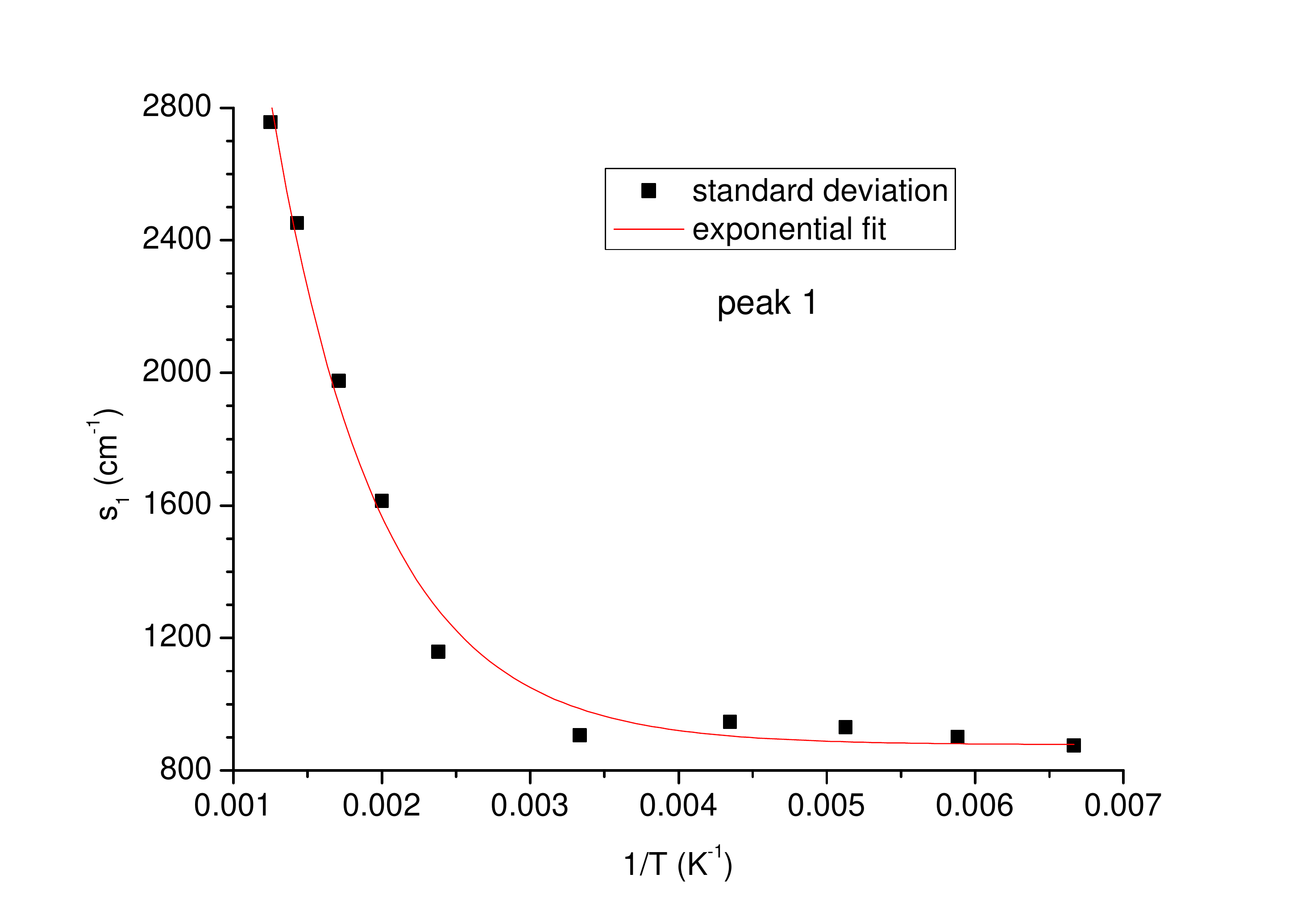}
\caption{\textit{Top:} Analytical representation of the first peak of $\sigma_{cont}(\lambda, T)$ with a gaussian function, $\sigma_1(\lambda,T)$, centred at 88574 cm$^{-1}$ for temperatures between 150 and 800 K. \textit{Bottom:} Variation of the standard deviation $s_1(T)$ of the first gaussian function as a function of 1/T. The variation can be fitted by a decreasing exponential function.}
\label{fig:pic1}
\end{figure}
To determine the parameterization, we proceeded in two steps. First, we fitted for each temperature the continuum of absorption with the sum of three gaussian functions. These fits are called hereafter "individual fits". Then, we traced the parameters (position centres, amplitudes, FWHMs of the three gaussian functions) of these "individual fits" in order to determine their variations with the temperature.

For the three gaussians, the positions of the centres ($\nu_{ci}$) do not vary with the temperature and are equal to $\nu_{c1} = 88574$  cm$^{-1}$ ($\sim$112.9 nm), $\nu_{c2} = 76000$  cm$^{-1}$ ($\sim$131.58 nm), and $\nu_{c3} = 68000$  cm$^{-1}$ ($\sim$147.06 nm). These values are quite close to the positions of the electronic states $1^1\Sigma^+_u$, $1^1\Pi_g$, and $1^1\Delta_u$, respectively \citep[and references therein]{cossart1987high,grebenshchikov2013photodissociation}.
\begin{figure}
\centering
\includegraphics[angle=0,width=\columnwidth]{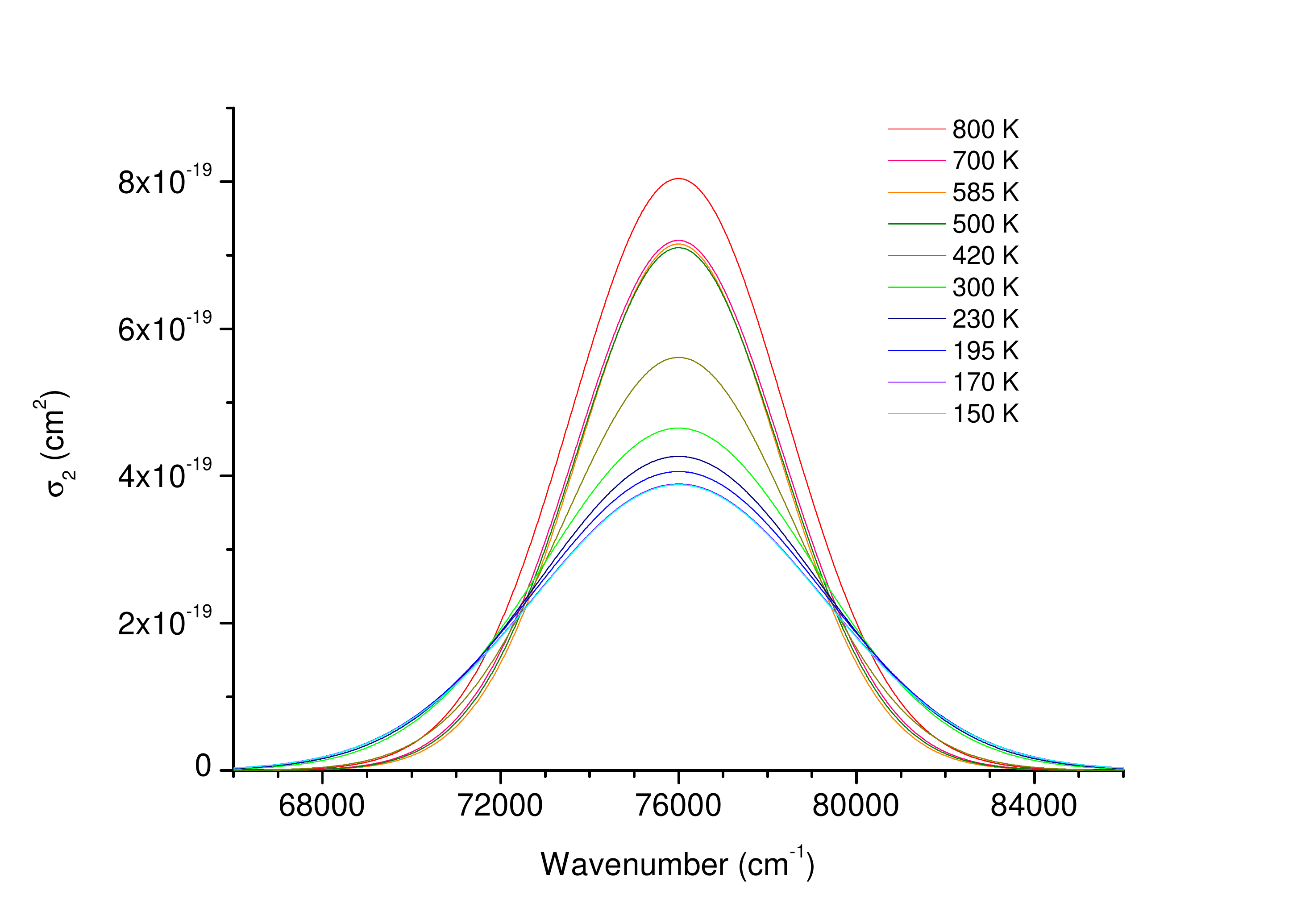}
\includegraphics[angle=0,width=\columnwidth]{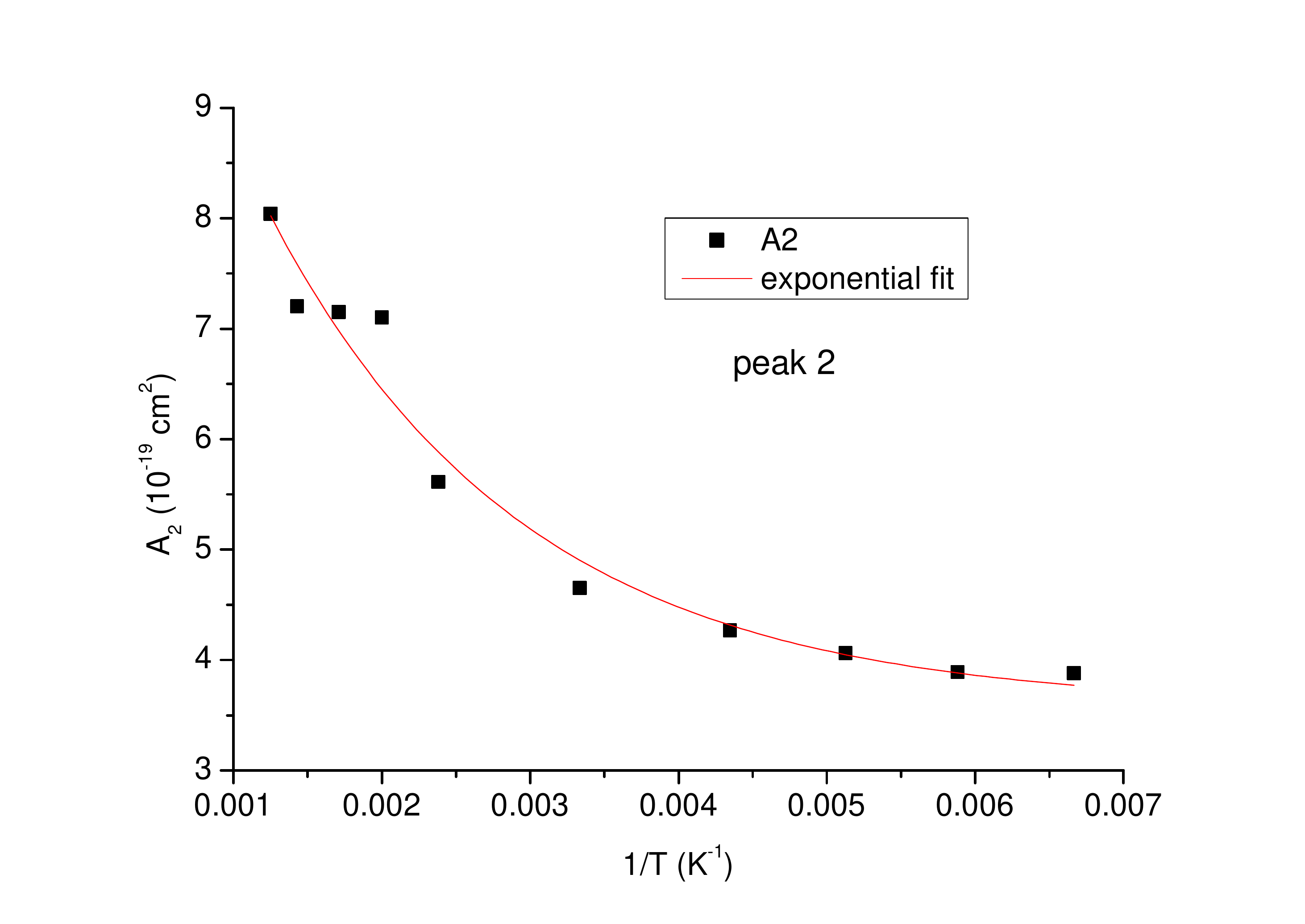}
\includegraphics[angle=0,width=\columnwidth]{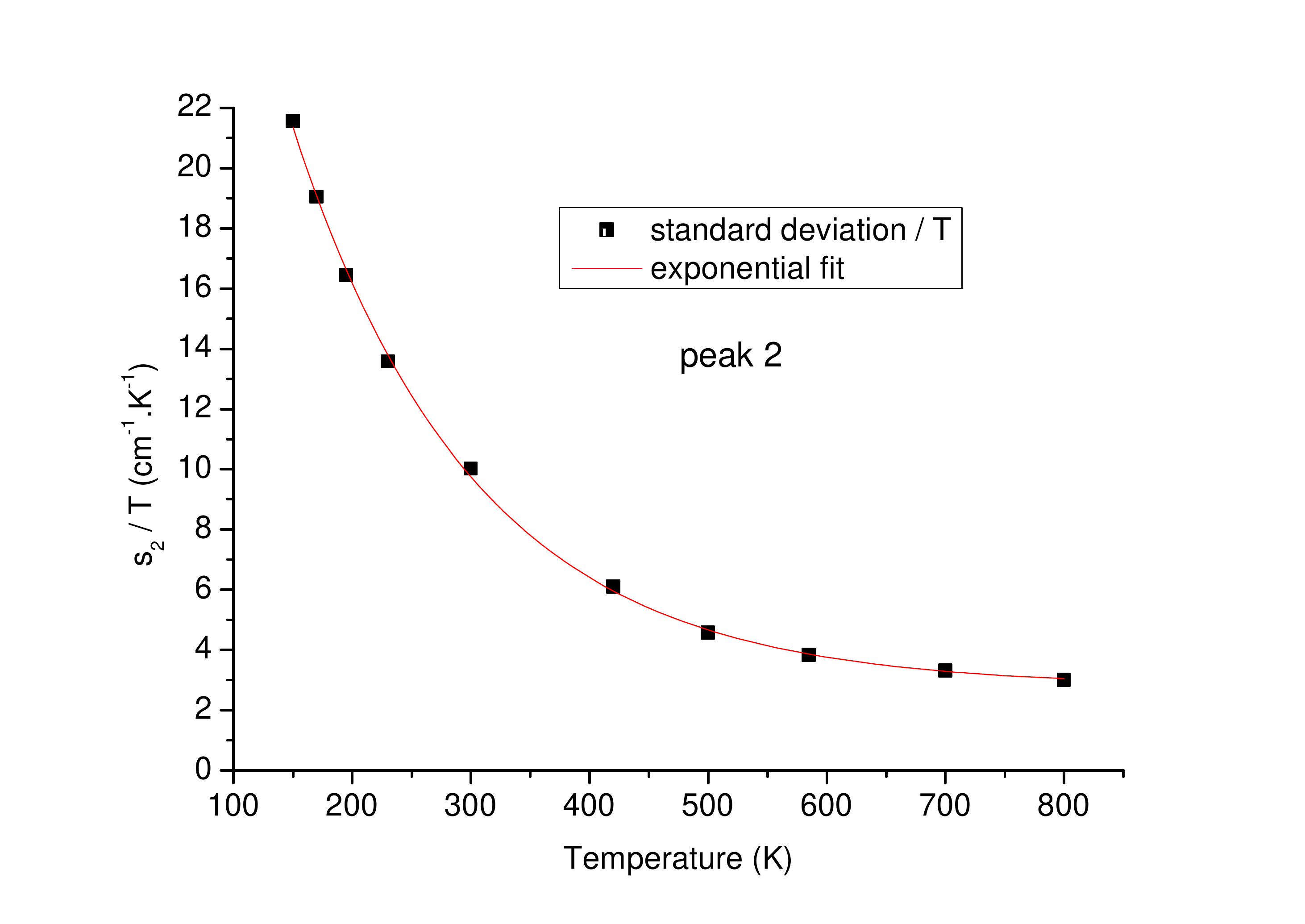}
\caption{\textit{Top:} Analytical representation of the second peak of $\sigma_{cont}(\lambda, T)$ with a gaussian function, $\sigma_2(\lambda,T)$, centred at 76000 cm$^{-1}$ for temperatures between 150 and 800 K. \textit{Middle:} Variation of the amplitude $A_2(T)$ of the second gaussian function as a function of 1/T. The variation can be fitted by a decreasing exponential function.
\textit{Bottom:} Variation of the standard deviation $s_2(T)/T$ of the second gaussian function as a function of the temperature. The variation can be fitted by a decreasing exponential function.}
\label{fig:pic2}
\end{figure}
\begin{figure}
\centering
\includegraphics[angle=0,width=\columnwidth]{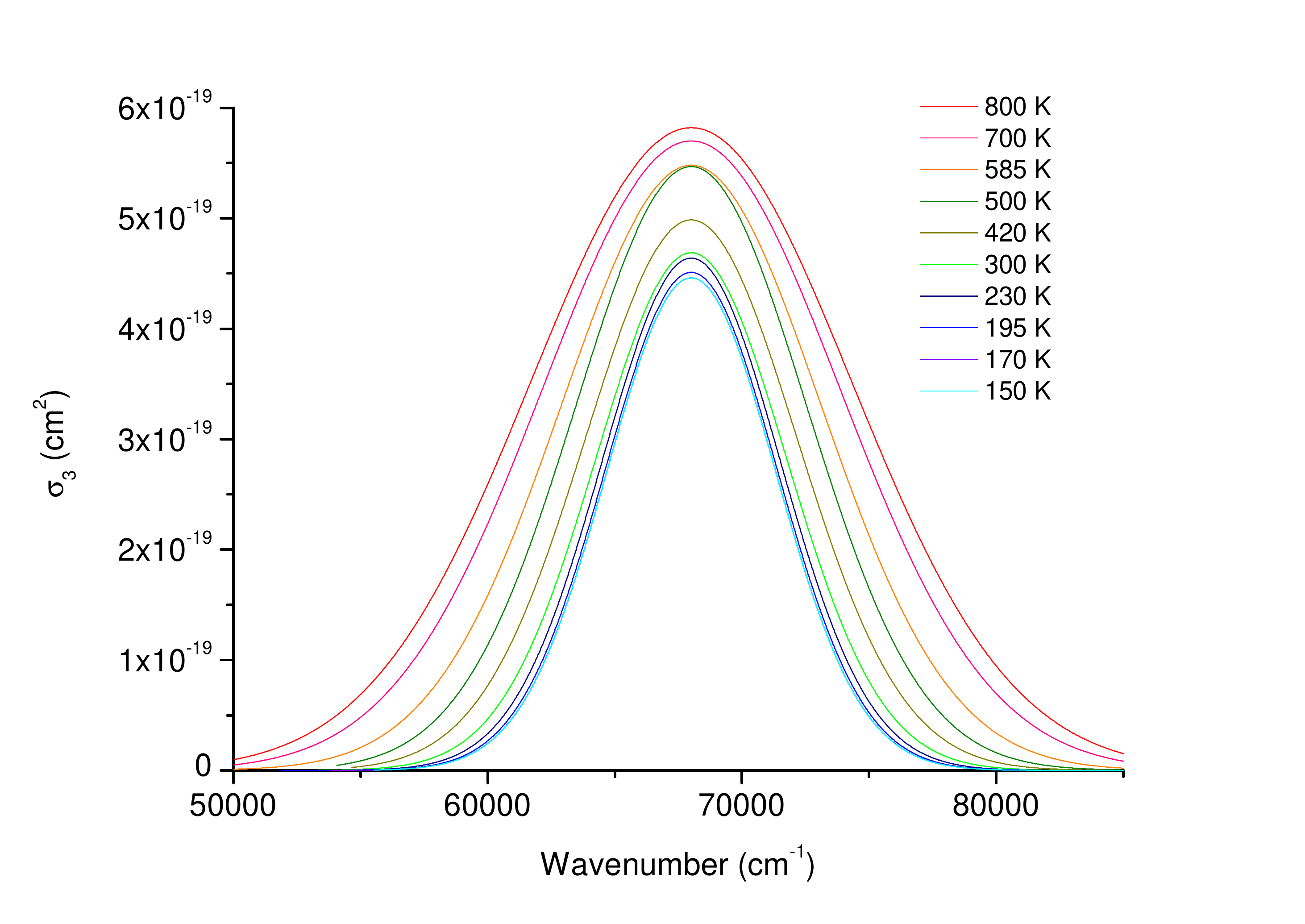}
\includegraphics[angle=0,width=\columnwidth]{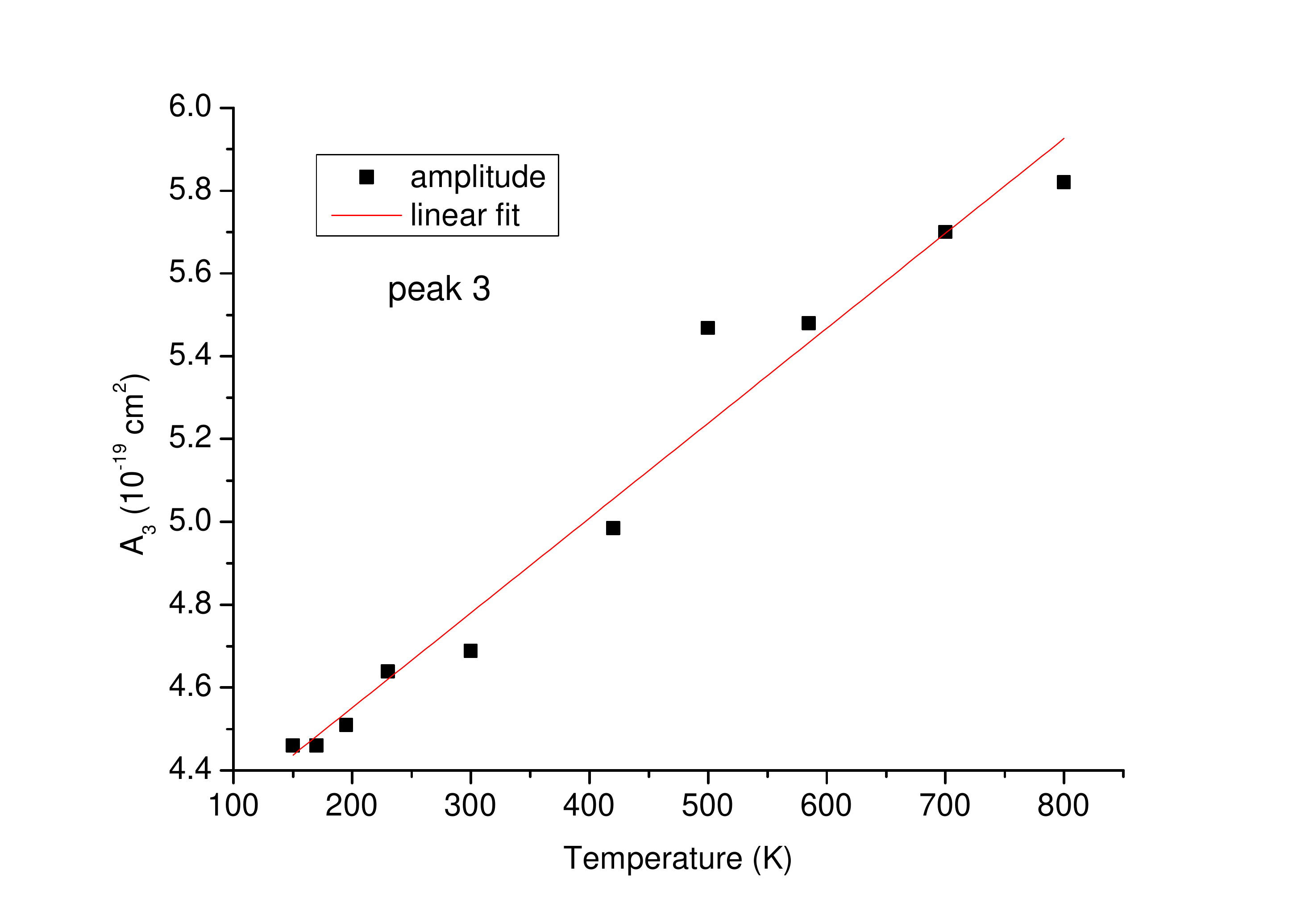}
\includegraphics[angle=0,width=\columnwidth]{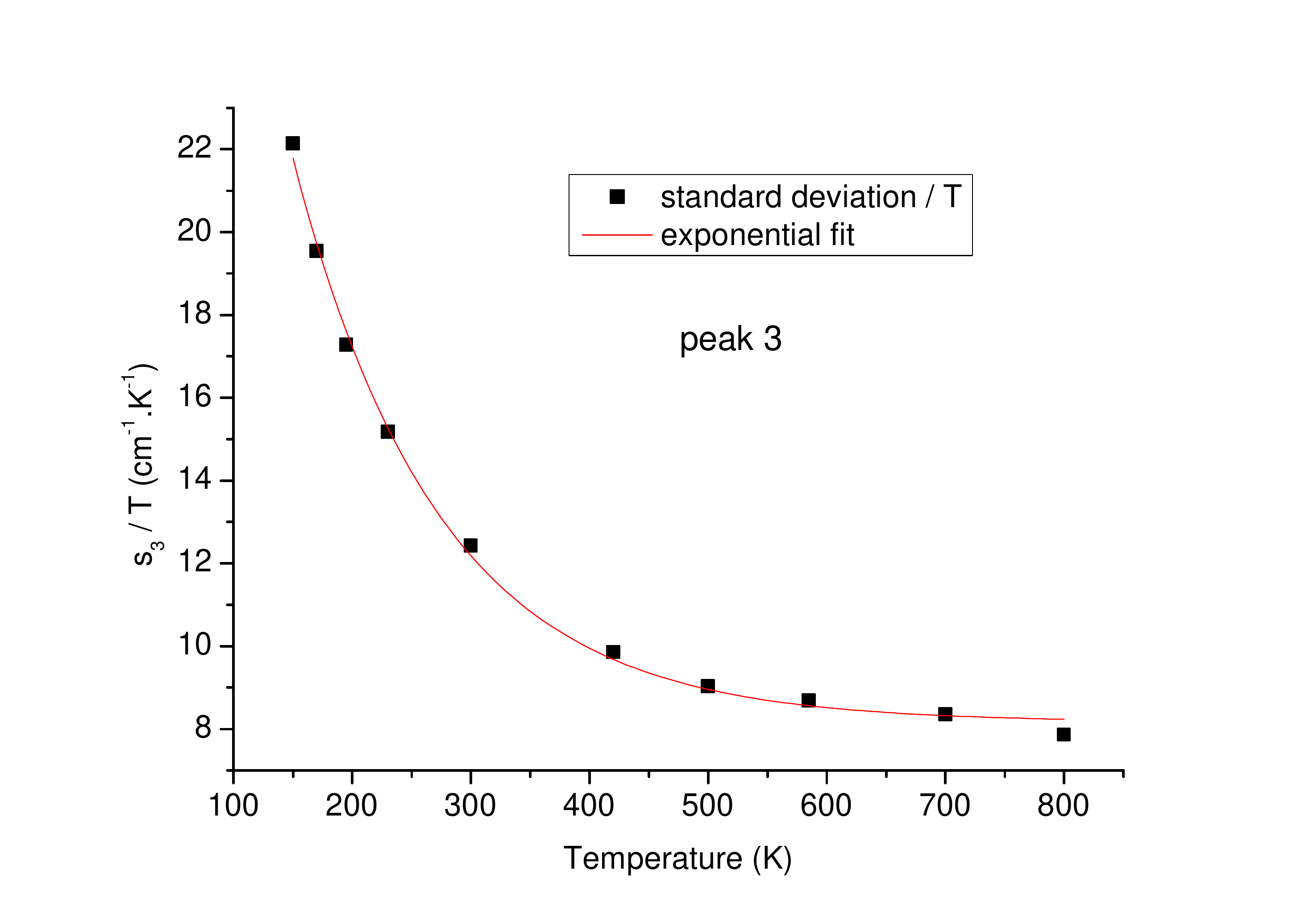}
\caption{\textit{Top:} Analytical representation of the third peak of $\sigma_{cont}(\lambda, T)$ with a gaussian function, $\sigma_3(\lambda,T)$, centred at 68000 cm$^{-1}$ for temperatures between 150 and 800 K. \textit{Middle:}  Variation of the amplitude $A_3$ of the third gaussian function as a function of the temperature. The variation can be fitted by a linear function. \textit{Bottom:} Variation of the standard deviation $s_3(T)/T$ of the third gaussian function as a function of the temperature. The variation can be fitted by a decreasing exponential function.}
\label{fig:pic3}
\end{figure}
For the first peak, the amplitude was fixed at $A_1=5\times10^{-20}$cm$^2$ for all the temperatures. Consequently, only the FWHM $s_1(T)$ depends on the temperature. We determined that the variation of $s_1(T)$ could be fitted by the decreasing exponential function (see Fig.~\ref{fig:pic1}) :
\begin{equation}\label{eq:s1}
s_1(T)= 877.36 + 10947.81 \times exp\left(-\frac{1382.63}{T}\right)
\end{equation}

For the two other gaussians, the amplitudes $A_2(T)$ and $A_3(T)$ vary with the temperature. As can been seen in Figs.~\ref{fig:pic2} and \ref{fig:pic3}, these variations can be fitted with a decreasing exponential function (Eq. \ref{eq:A2}) and a linear function (Eq. \ref{eq:A3}), respectively. 
\begin{equation}\label{eq:A2}
A_2(T)= 3.58 + 9.18 \times exp\left(-\frac{580.92}{T}\right)
\end{equation}
\begin{equation}\label{eq:A3}
A_3(T)= 4.09 + 0.0022 \times T
\end{equation}
Concerning the FWHMs, their variations with respect to temperature are very well fitted by decreasing exponential functions (Eqs. \ref{eq:s2} and \ref{eq:s3}).
\begin{equation}\label{eq:s2}
s_2(T)= T \times ( 2.78 + 49.52 \times exp(-0.00654 \times T) )\\
\end{equation}
\begin{equation}\label{eq:s3}
s_3(T)= T \times ( 8.17 + 46.12 \times exp(-0.00813 \times T) )
\end{equation}

\begin{figure}[!htb]
\centering
\includegraphics[angle=0,width=\columnwidth]{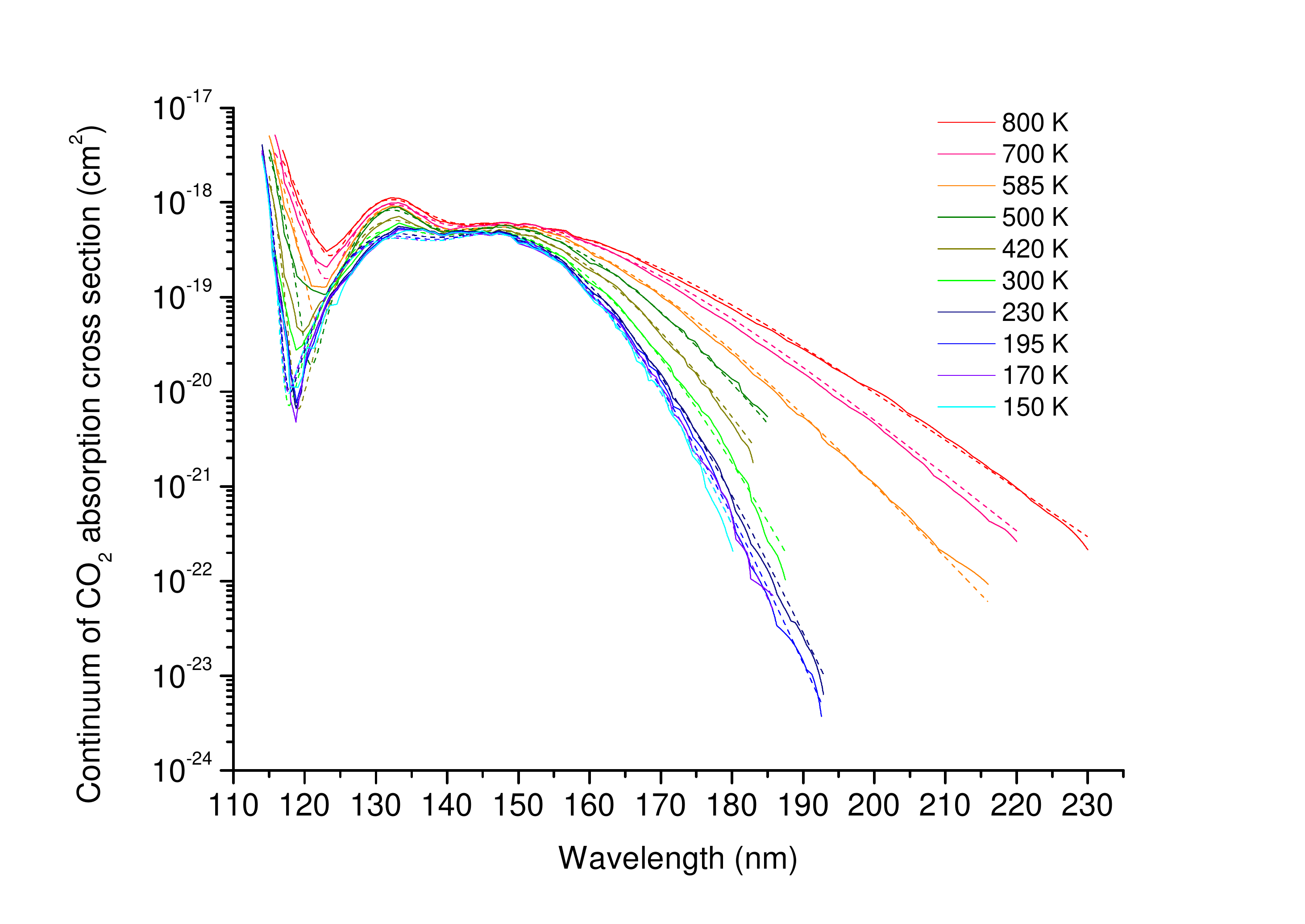}
\caption{Continuum of CO$_2$ absorption cross section (cm$^2$) measured between 150~K and 800 K (full lines). The continuum can be fitted by a sum of three gaussian functions (dashed lines).}
\label{fig:continu_fit}
\end{figure}

From Eqs.~\ref{eq:sigmatot} and \ref{eq:sigmai}, these parameterisations allow us to calculate the continuum of the absorption cross section of CO$_2$ at any wavelength and any temperature in the ranges (115--230 nm) and (150--800 K).
The continuums determined from the experimental data and from the analytical formulation are represented in Fig.~\ref{fig:continu_fit} for all the temperatures between 150 and 800 K. One can see that the parameterization gives very good results, close to the experimental data. On average, the mean deviation between the experimental data and the parameterization is about 15\%.

%

\subsubsection{Extrapolation at temperatures $>$ 800~K}
\begin{figure}[!ht]
\centering
\includegraphics[angle=0,width=\columnwidth]{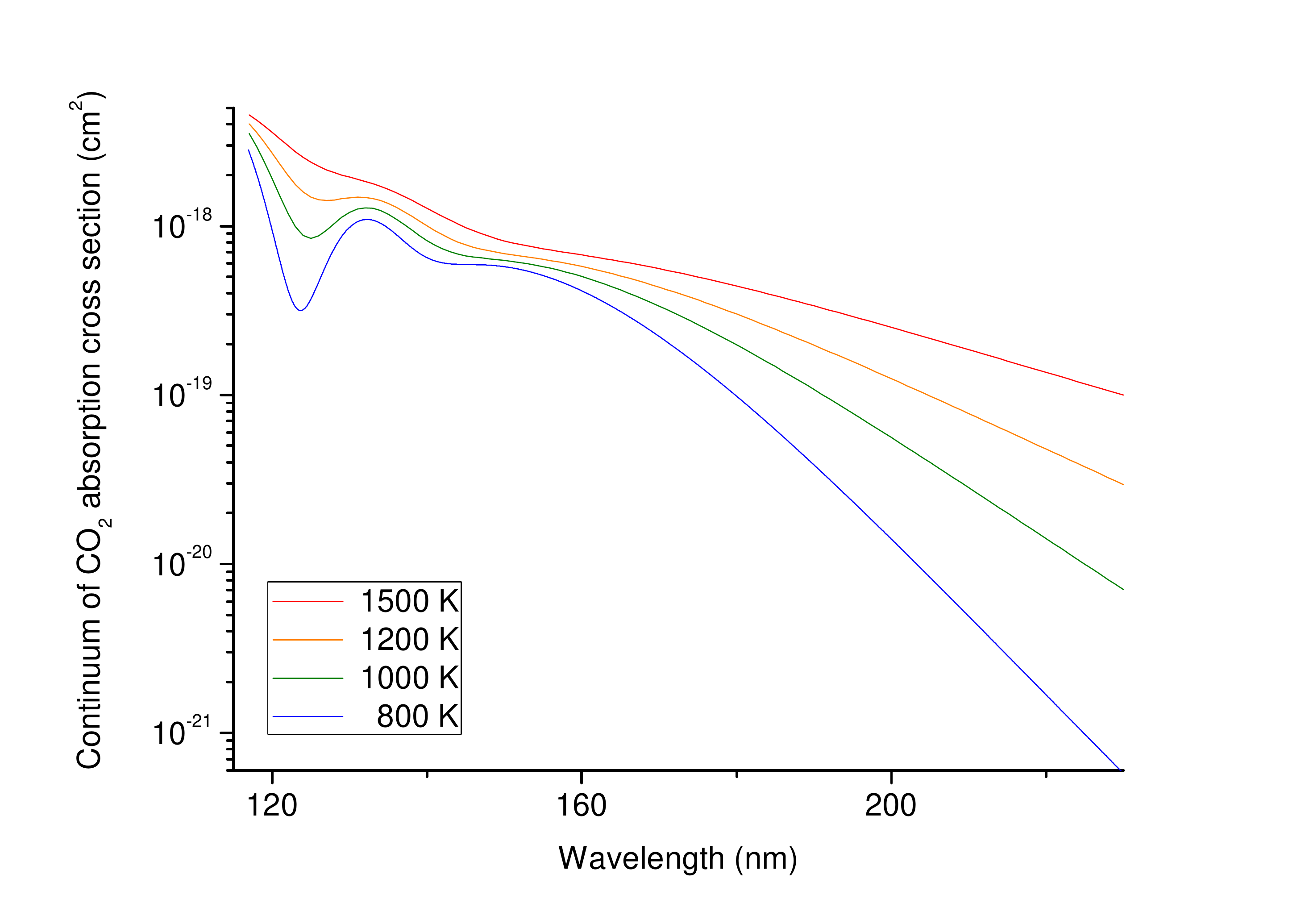}
\caption{Continuum of CO$_2$ absorption cross section extrapolated to 1000~K (\textit{green}), 1200~K (\textit{orange}), and 1500~K (\textit{red}). $\sigma_{cont}(\lambda, 800)$ (\textit{blue}) is shown for comparison.}
\label{fig:extrapol}
\end{figure}

We used Eq.~\ref{eq:sigmatot} to extrapolate absorption data to higher temperatures (Fig.~\ref{fig:extrapol}). One can see that as long as the temperature increases, the wings of the gaussian functions broaden and the troughs between the three peaks disappear. We compared our results with the parameterisations published by \cite{Schulz2002} and \cite{oehlschlaeger2004ultraviolet} for the wavelengths longer than 190 nm, but found a disagreement. For one given temperature, the absorption cross sections determined using their coefficients is lower than our results. To obtain an absorption cross section compatible with our results with their parameterisation, the temperature has to be increased by a factor $\sim$1.5. This conclusion is similar to what we found in \cite{venot2013high}. The discrepancy between our results and theirs can have several origins. It might be due to an overestimation of the temperature or an underestimation of the absorbance in \cite{Schulz2002} and \cite{oehlschlaeger2004ultraviolet}. It has also been pointed out by \cite{oehlschlaeger2004ultraviolet} that the average mole fraction of CO$_2$ during the experiments of \cite{Schulz2002} has been calculated with a kinetic mechanism (i.e. GRIMech 3.0) that overestimated the thermal decomposition of CO$_2$, leading to too low abundances of carbon dioxide and thus too large absorption cross-sections.
\cite{Grebenshchikov2016}, who also noticed a difference in the temperature scale with \cite{oehlschlaeger2004ultraviolet} (by $\sim$ 300K), suggested that this discrepancy could find its origin in some assumptions made in his current model, but also in the development of local vibrational temperature regions in experiments of \cite{oehlschlaeger2004ultraviolet}. These local regions could have an impact on the absorption cross sections. The vibrational relaxation time, which has not been measured, might be different in our experiments and in \cite{oehlschlaeger2004ultraviolet}.


\section{Application to warm exoplanet atmospheres}\label{sec:application}

\subsection{Model}

\subsubsection{1D neutral chemical model}
We used our 1D thermo-photochemical model to study the effect of the absorption cross sections of CO$_2$ at high temperature on the chemical composition of exoplanetary atmospheres. Our goal here was not to model a real planet and predict future observations, but only to quantify the effect of the new data absorption cross section of carbon dioxide on the atmospheric composition predicted by chemical models. Our 1D time-dependent model is very well adapted to the study of warm atmospheres thanks to the chemical scheme it uses. This scheme has been validated experimentally in large ranges of pressure and temperature ([0.01--100] bars and [300--2500] K). It contains 105 species, made of H, C, O, and N and linked by 1920 neutral-neutral reactions. To this validated chemical scheme, we added 55 photolysis reactions. Because our core scheme is a non-optimized network (i.e. the reaction rate coefficients are those recommended by kinetics databases and have not been modified to fit experiments), the addition of these photolyses reactions does not call into question the validation of the core scheme (see \citealt[and references therein]{venot2012}). An experimental validation of the whole scheme with photolysis processes would be ideal but represents a challenge given the lack of knowledge to date on high-temperature photolysis data.
We used solar C/N/O relative abundances but increased the metallicity by a factor 100 compared to the solar metallicity \citep{asp2009}, as high metallicity is probable in warm gaseous atmospheres \citep[e.g.][]{fortney2013, moses2013, agundez2014}. The consequence is an increase of the abundance of CO$_2$, compared to solar metallicity atmosphere, by approximately 4 orders of magnitude, resulting, for instance at 1 mbar, in molar fractions of CO$_2$ ($y_{CO_2}$) of 6$\times$10$^{-3}$  and 6$\times$10$^{-4}$ for the two atmospheres studied here, at  800~K and 1500~K respectively(see Sect.~\ref{sec:thermal}).

\subsubsection{Thermal profiles}\label{sec:thermal}
\begin{figure}[!htb]
\centering
\includegraphics[angle=0,width=\columnwidth]{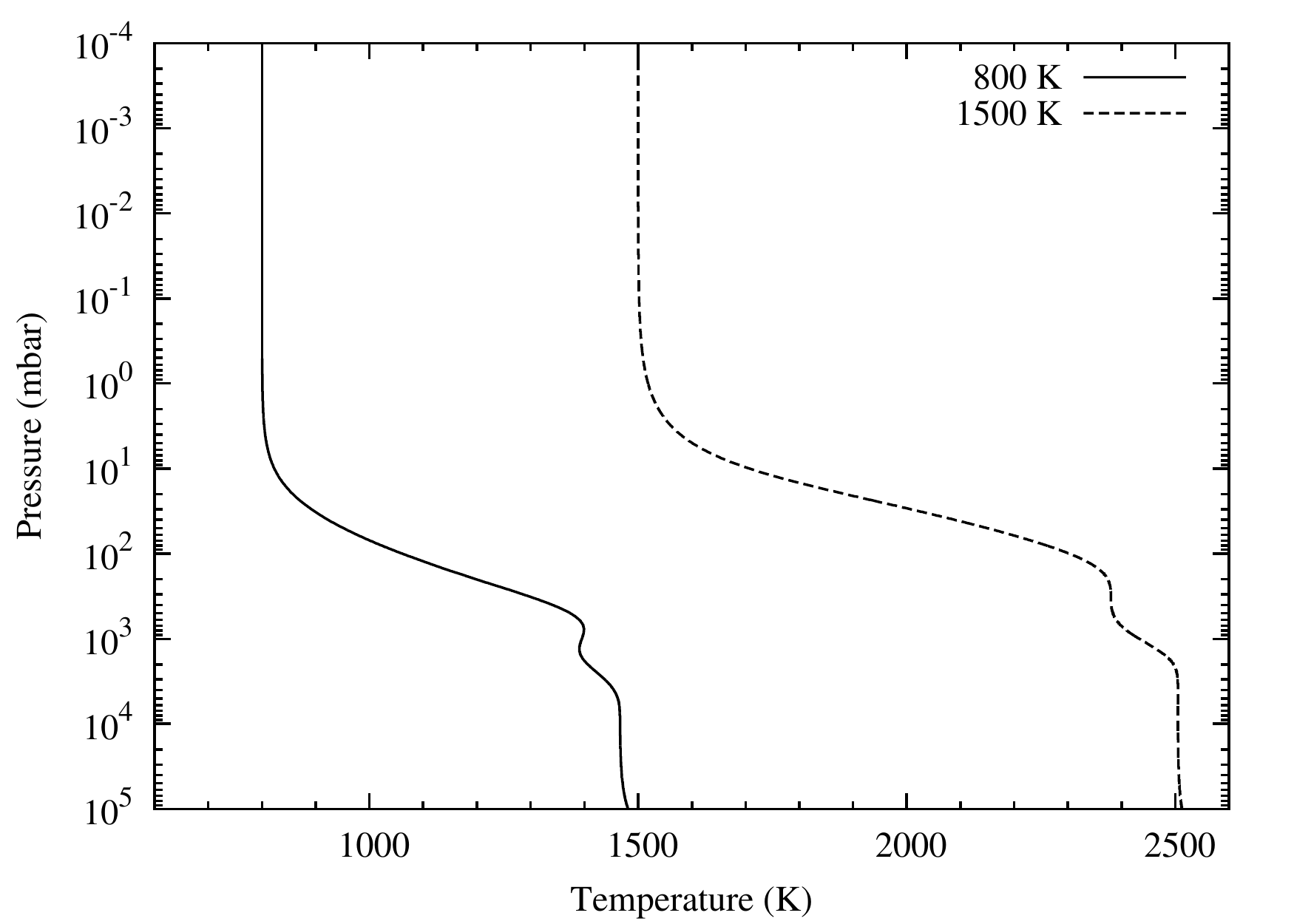}
\caption{Thermal profiles used in our thermo-photochemical model to study the impact of CO$_2$ high-temperature absorption cross sections.}
\label{fig:PTprofile}
\end{figure}

We modelled two warm Neptunes with physical characteristics (mass and radius) similar to those of GJ~436b \citep{southworth2010homogeneous} and with equilibrium temperatures of 1091~K and 2043~K, which lead to temperatures of 800~K and 1500~K, respectively, in the upper part of the atmosphere, above 10 mbars (Fig.\ref{fig:PTprofile}).
To construct the thermal profiles, we used the analytical model of \cite{Parmentier2015}, with opacities from \cite{valencia2013bulk} and other opacities representing stellar light absorption and non-grey thermal effects adjusted to reproduce numerical models similar to the ones of \cite{fortney2005}. We did not include TiO and VO in the calculation of the radiative transfer. Note that at equilibrium temperatures less than $\sim 1800-1900$~K, the low gaseous abundance of TiO and VO, due to their cold trap in the deep atmospheric layers, has no effect on the thermal structure \citep{fortney2008, Parmentier2016}.
The chemical composition was assumed to be solar, which is not consistent with the atmospheric metallicity (100 $\times$ solar) we used in our modelling. However, as the purpose of this application is not to reproduce a real planet and its observations but simply to see the effect of the different absorption cross sections of carbon dioxide, we did not try to calculate more realistic thermal profiles. Furthermore, a thermal profile consistent with a high metallicity would have a higher temperature in the deep atmosphere but the temperature in the upper atmosphere (for $P <$ 1 mbar) would remain the same \citep{lewis2010, agundez2014}.

\subsubsection{Stellar irradiation}
We chose a star emitting a high flux in the range (130--230 nm): HD~128167 (stellar type F2V). The stellar spectrum was constructed with observational data of HD~128167  for $\lambda$ $\in$ (115--900 nm) \citep{segura2003ozone}. For $\lambda$ $\in$ (1--114 nm) we used the solar spectrum \citep{thuillier2004solar} that we scaled to the radius and temperature of HD~128167 (i.e. $R_{\star}=1.434~R_{\odot}$ and $T_{\star}=6723$~K), in order to have a luminosity consistent with the physical parameters of the star. 
To obtain a stellar irradiation consistent with the equilibrium temperatures 1091~K and 2043~K, we set the semi-major axis to 0.126 AU and 0.036 AU, respectively.
In \cite{venot2013high}, we used three different stars (F, G, and M) and observed more variations of chemical abundances with the F star than with the two other ones. Thus, here we limited our present study to this spectral type only.

\subsubsection{Eddy diffusion coefficient}
For the vertical mixing, we considered a constant eddy coefficient $K_{zz}=10^8$~cm$^2$s$^{-1}$. This value is often used in 1D photochemical models and corresponds to an average value deduced from 3D Global Circulation Models (GCMs) when one multiplies the vertical scale height by the root mean square of the vertical velocity. To date, no consensus exists on the best method to express the vertical mixing in 1D models and \cite{par2013} showed that the eddy diffusion coefficient can vary by two orders of magnitude depending on the method chosen.

\subsubsection{CO$_2$ absorption cross sections}
We modelled the warm Neptunes described above using absorption cross sections of CO$_2$ presented in this paper. For the planet with a temperature of 800 K in the upper atmosphere, we used (a) the absorption cross section at ambient temperature: $\sigma_{CO_2}(\lambda, 300)$, (b) the absorption cross section at 800 K: $\sigma_{CO_2}(\lambda, 800)$, and (c) the continuum of absorption extrapolated with Eq.~\ref{eq:sigmatot} at 800~K: $\sigma_{cont}(\lambda, 800)$.
For the warmer planet, with a temperature of 1500 K at the top of the atmosphere, we used (a) $\sigma_{CO_2}(\lambda, 300)$ and (b) the continuum of absorption calculated with Eq.~\ref{eq:sigmatot} at 1500~K: $\sigma_{cont}(\lambda, 1500)$.
All the absorption cross sections have been binned to a resolution $\Delta \lambda = 1$ nm in order to optimized the computational time. Doing so, the fine structure of the absorption (visible in Fig.\ref{fig:crosssection}) is smoothed.
Stellar absorption due to the other molecules was calculated using data at the highest temperature available in the literature, which is often between 300 and 400 K. The references concerning the absorption cross sections and the quantum yields of photodissociation can be found in \cite{venot2012} and \cite{dob2014}. The use of these mid-temperature data is the main source of uncertainty in our atmospheric modelling. One has to keep in mind that the atmospheric compositions presented in Figs. \ref{fig:abundances} and \ref{fig:abundances_1500K} might be different in reality. Here, our main interest is to show the influence of the hot CO$_2$ absorption cross sections.

\subsubsection{CO$_2$ photodissociation rates}
Under stellar irradiation, carbon dioxide has two routes to photodissociate :
\begin{align}
\mathrm{CO_2 + h\nu \longrightarrow CO + O(^3P) \qquad J_1(z, T)}\\
\mathrm{CO_2 + h\nu \longrightarrow CO + O(^1D) \qquad J_2 (z, T)}
\end{align}
The route favoured depends on the energy of the photons, according to the quantum yields presented in Table~\ref{tab:q} \citep{huebner1992solar}. These data have been measured at room temperature, but in the absence of data at higher temperature, we assume the same values at 800 K and 1500 K.

The loss rate of CO$_2$, due to photolysis only, is given by:
\begin{equation}
L_{CO_2}^{phot} = \sum_{k=1,2} - J_k(z,T) n_{CO_2},
\end{equation}
where $n_{CO_2}$ is the density of CO$_2$ (cm$^{-3}$) and $J_k(z,T)$ are the photodissociation rates of CO$_2$ (s$^{-1}$). The photodissociation rate depends on the absorption cross section of CO$_2$, $\sigma_{CO_2}(\lambda, T)$ (cm$^2$), the actinic photon flux, $F(\lambda, z, T)$ (cm$^{-2}$.s$^{-1}$.nm$^{-1}$), and the quantum yield corresponding to the route $k$, $q_k(\lambda)$, through the equation:
\begin{equation}\label{eq:J}
J_k(z, T) = \int_{\lambda_1}^{\lambda_2} \sigma_{CO_2}(\lambda, T) F(\lambda, z, T) q_k(\lambda) d\lambda,
\end{equation}
where $[\lambda_1;\lambda_2]$ is the spectral range on which CO$_2$ absorbs the UV flux.

Carbon dioxide interacting chemically with other species in the atmosphere, the global variation of CO$_2$ abundance over time is determined by the continuity equation:
\begin{equation}\label{eq:continuite}
\frac{\partial n_{i}}{\partial t} = P_{i} - L_{i} - div({\Phi_{i}}\overrightarrow{e_z})
\end{equation}
where for a species $i$, $P_{i}$ is its the total production rate ($\mathrm{cm^{-3}.s^{-1}}$), $L_{i}$ its total loss rate ($\mathrm{cm^{-3}.s^{-1}}$), and $\Phi_{i}$ its vertical flux ($\mathrm{cm^{-2}.s^{-1}}$), which follows the diffusion equation.

This continuity equation governs the evolution of abundances of every molecules present in the atmosphere. Thus, all species are linked through a complex non-linear system of differential equations.

\begin{table}[!h]
\begin{center}\begin{tabular}{ll}
\hline
\hline
Quantum yield & Values [wavelength range] \\
\hline
$q_1(\lambda)$ & 1 [167-227] ; 0 elsewhere \\
$q_2(\lambda)$ & variable [50-107] ; 1 [108-166] ; 0 elsewhere \\
\hline
\end{tabular}\end{center}
\caption{Quantum yields for the photodissociations of carbon dioxide.}\label{tab:q}
\end{table}

\subsubsection{Synthetic transmission spectra}

For the different atmospheric compositions determined with the 1D kinetic model, we computed synthetic infrared transmission spectra in the wavelength range  $0.4-25$\,$\mu$m using the code Tau-REx \citep{waldmann2015, waldmann2015b} in forward mode, which allows variable pressure-dependent temperature profiles. A discussion on the effect of isothermal vs. non-isothermal profiles can be found in \cite{rocchetto2016}. The infrared absorption cross sections of the absorbing species were computed using the linelists from ExoMol \citep{Tennyson2012}, HITRAN \citep{Rothman2009,Rothman2013} and HITEMP \citep{Rothman2010}. The absorbing species considered are C$_2$H$_2$, CO$_2$, HCN, NH$_3$, CH$_4$, CO, and H$_2$O. We considered collision-induced absorption of H$_2$-H$_2$ and  H$_2$-He \citep{Rothman2009,Rothman2013} and assumed that the atmosphere is cloud-free. The spectrum shown was binned to a spectral resolution $R=\lambda/\Delta \lambda$, constant in wavelength, of 300.

\subsection{Results}

\subsubsection{Photodissociation rates}\label{sec:rates}
\begin{figure}[!htb]
\centering
\includegraphics[angle=0,width=\columnwidth]{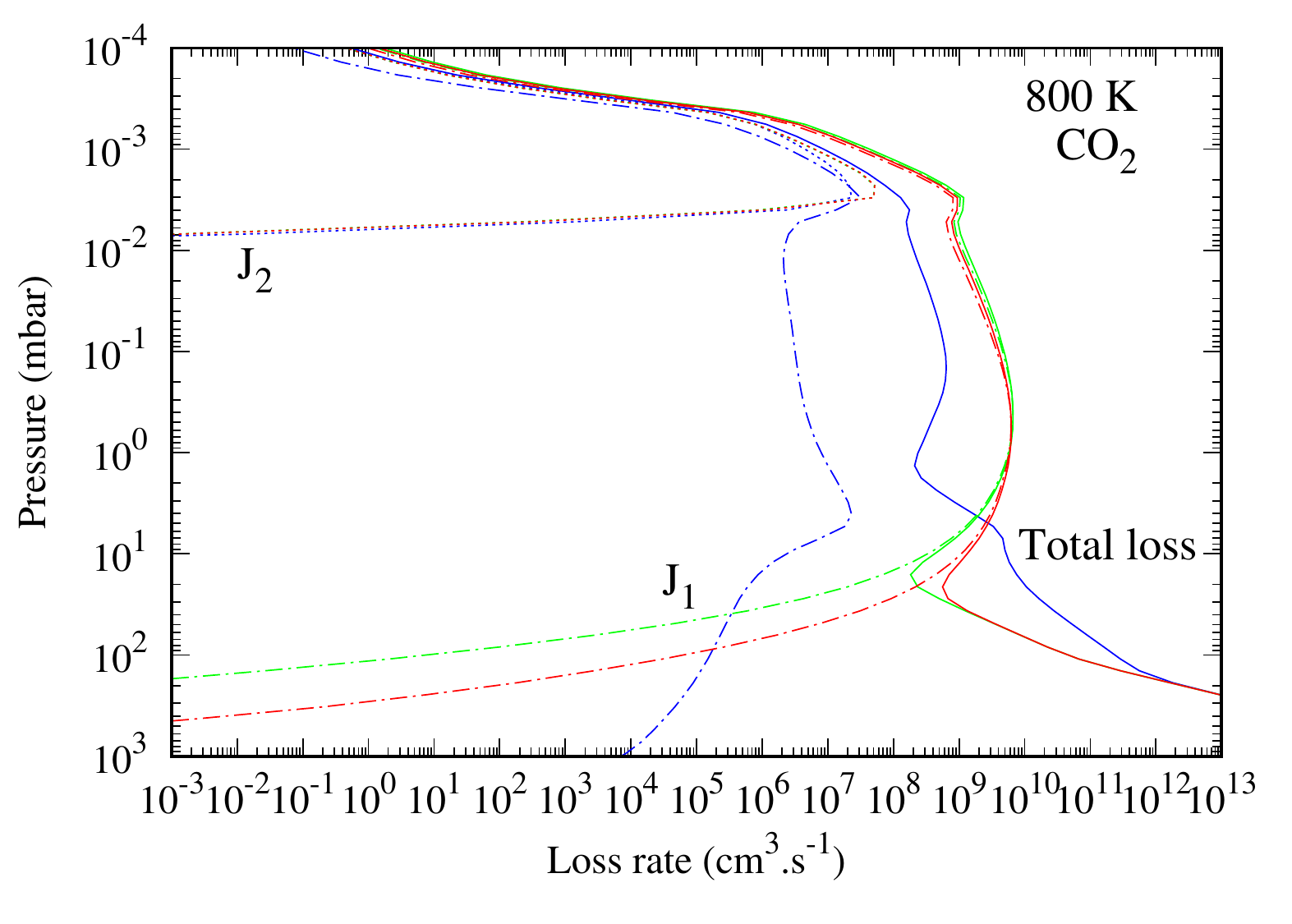}
\includegraphics[angle=0,width=\columnwidth]{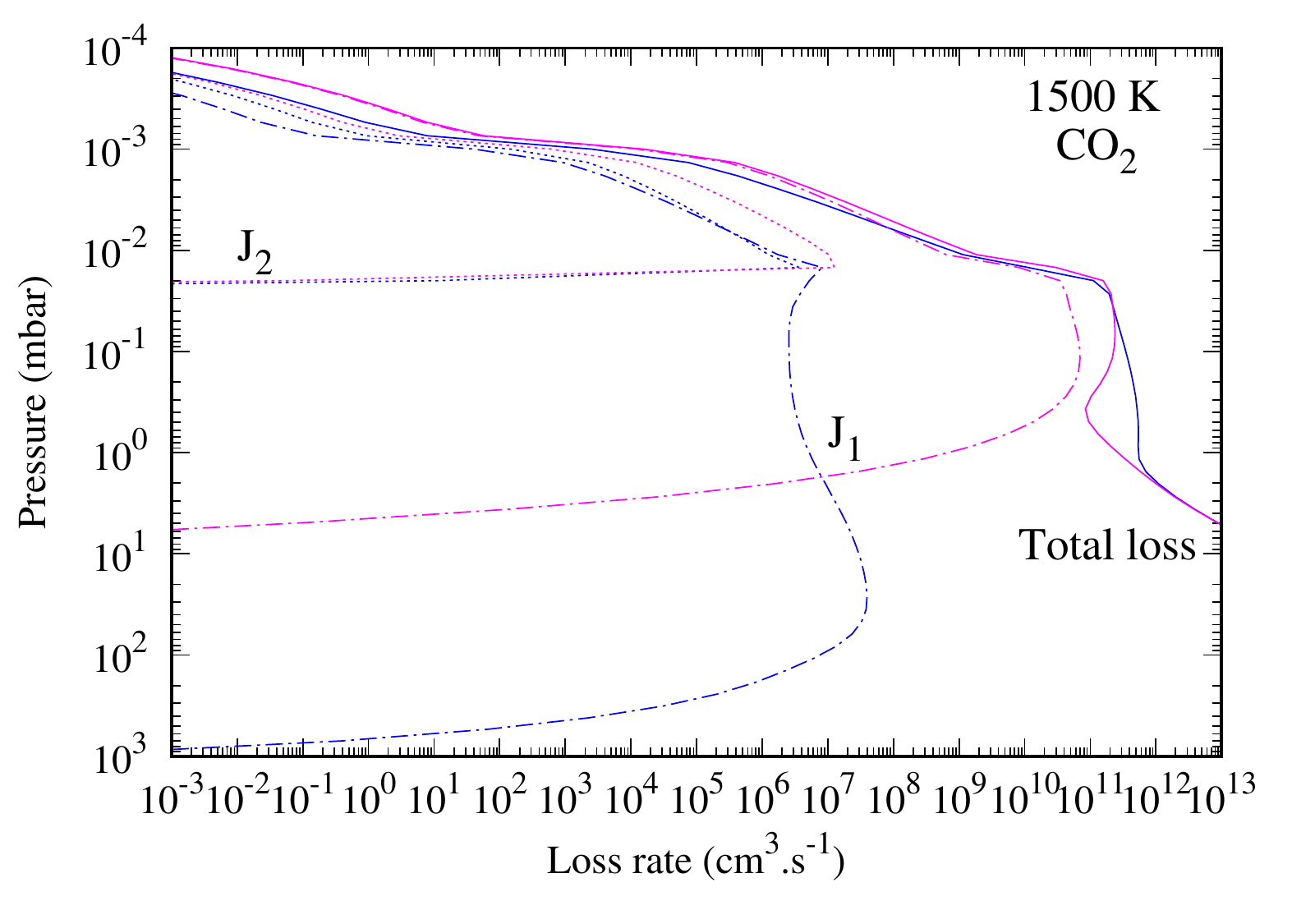}
\caption{For CO$_2$, total loss rates (\textit{full line}) and loss rates due to photolysis J$_1$ (\textit{dotted-dashed line}) and J$_2$ (\textit{dotted line}) in the two atmospheric models: "800~K" (\textit{top}) and "1500~K" (\textit{bottom}). Colors correspond to the different absorption cross section used: $\sigma_{CO_2}(\lambda, 300)$ (\textit{blue}), $\sigma_{CO_2}(\lambda, 800)$ (\textit{red}), $\sigma_{cont}(\lambda, 800)$ (\textit{green}), and $\sigma_{cont}(\lambda, 1500)$ (\textit{pink}).}
\label{fig:photo_rates}
\end{figure}

\begin{figure}[!htb]
\centering
\includegraphics[angle=0,width=\columnwidth]{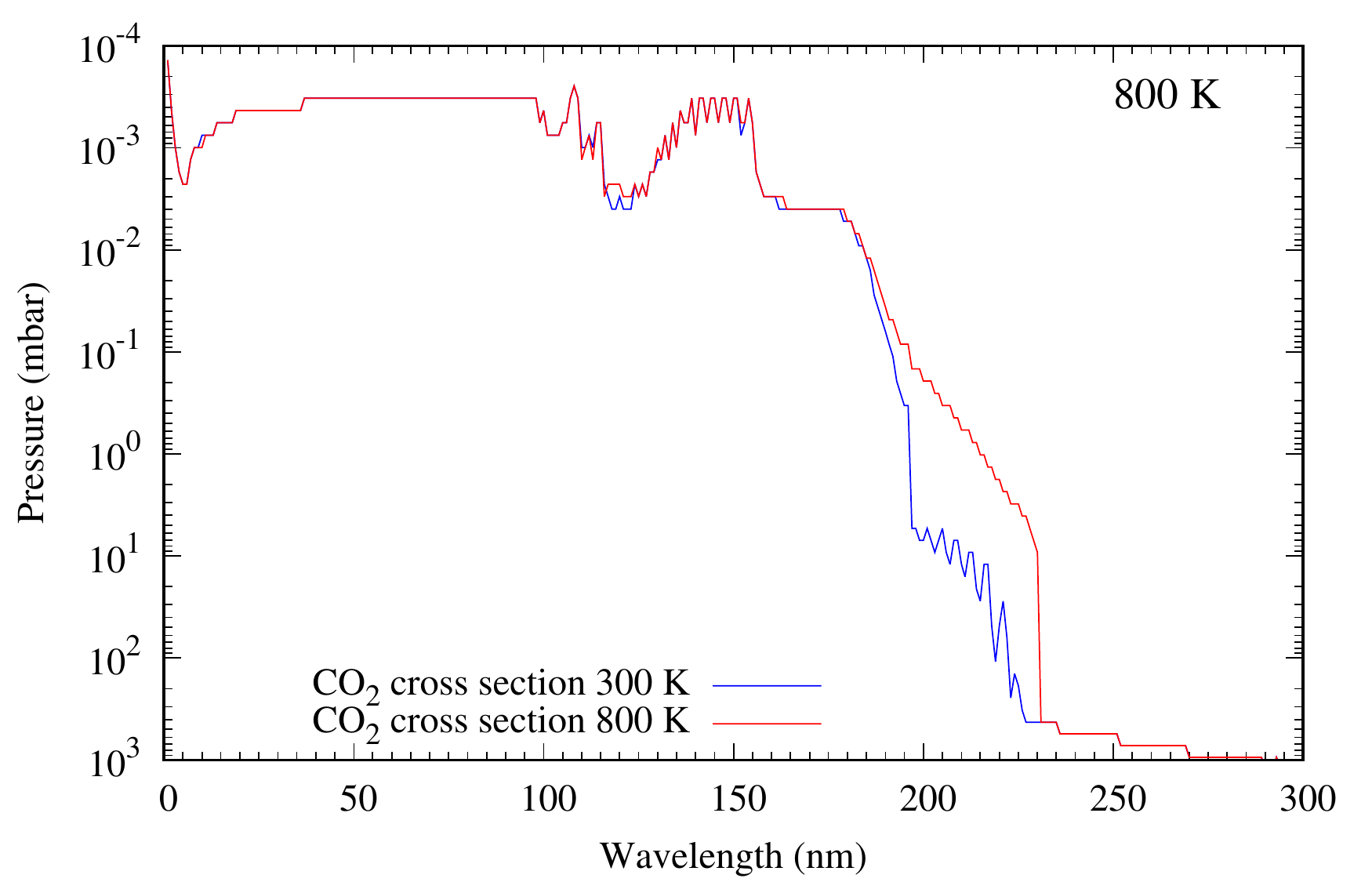}
\caption{Penetration of the stellar flux  in the atmosphere labelled "800~K" when using $\sigma_{CO_2}(\lambda, 300)$ (\textit{blue}) and $\sigma_{CO_2}(\lambda, 800)$ (\textit{red}). It represents the level where the optical depth $\tau$=1 in function of wavelength.}
\label{fig:pene}
\end{figure}

We represented in Fig.~\ref{fig:photo_rates} the total loss rate of CO$_2$ as well as the loss rates due to the photodissocations J$_1$ and J$_2$ for our different atmospheric models.
Changing the absorption cross section of CO$_2$ modifies the total loss rate of CO$_2$, and the loss rates due to the photodissociations J$_1$ and J$_2$. When using $\sigma_{CO_2}(\lambda, 800)$, instead of $\sigma_{CO_2}(\lambda, 300)$, the loss rate of J$_2$ growths by a factor $\sim$2.5 at 2$\times$10$^{-3}$ mbar. However, the major change is observed for J$_1$, which has a loss rate multiplied by $\sim$1200 at 0.1 mbar, when using the warm absorption cross section. Figure \ref{fig:pene} shows that with $\sigma_{CO_2}(\lambda, 800)$, the stellar irradiation (for $\lambda <$ 230 nm) penetrates less deeply in the atmosphere because CO$_2$ absorbs more incoming flux with the warm absorption cross section than with the 300~K one. Thus, the loss rate of J$_1$ decreases for pressures higher than 20 mbar and is finally less important than in the model with $\sigma_{CO_2}(\lambda, 300)$ for P$>$100 mbar. However, at such pressures, the total loss rate of CO$_2$ is not due to the photodissociations anymore. It is at low pressures that the contribution of the photodissociations is important for the total loss rate. In the model using $\sigma_{CO_2}(\lambda, 800)$, the loss due to J$_1$ represents more than 90$\%$ of the total loss of CO$_2$ (for P$<$ 20 mbar), whereas this photodissociation has only a minor contribution to the total loss rate of CO$_2$ (0.5$\%$) when using $\sigma_{CO_2}(\lambda, 300)$. One can see that the total loss rate of CO$_2$ increases when $\sigma_{CO_2}(\lambda, 800)$ is used in the model. For instance, at 0.5 mbar, it has been multiplied by $\sim$18.

Comparison between the results obtained with the experimental data and our parameterization is satisfying. One can see that the differences on the loss rates are quite small between the models using $\sigma_{CO_2}(\lambda, 800)$ and $\sigma_{cont}(\lambda, 800)$. For the loss rate of J$_2$, the deviation is 3$\%$ at 2$\times10^{-3}$ mbar. For the one of J$_1$, the deviation is 3$\%$ at 1 mbar but increases as long as the pressure increases. It reaches $\sim$70$\%$ at 10 mbar and 100$\%$ at 100 mbar. As the photodissociation J$_1$ is a major actor in the destruction of CO$_2$, the total loss rates is thus also different. The maximum difference is 74$\%$ at 16 mbar. For higher pressures, the contribution of the photodissociations to the total loss of CO$_2$ is negligible and the total loss rate is the same whether we use $\sigma_{CO_2}(\lambda, 800)$ or $\sigma_{cont}(\lambda, 800)$.

Unlike the previous case around 800~K, when one compares the models using $\sigma_{CO_2}(\lambda, 1500)$ and $\sigma_{CO_2}(\lambda, 300)$, one notices that the total loss rate of CO$_2$ undergoes only slight modifications. The total loss rate of CO$_2$ is even lowered by a factor 6 at 0.37 mbar, because of loss rates of other chemical reactions, whose contributions on the total loss rate are not shown. Qualitatively, we observe the same behaviour than for the atmosphere at 800~K. In the upper atmosphere, the loss rates of J$_2$ and J$_1$ increase when the warmer absorption cross section is used. We notice that the loss rate of J$_1$ begins to decrease for lower pressures in the model using $\sigma_{CO_2}(\lambda, 1500)$ rather than $\sigma_{CO_2}(\lambda, 300)$. Quantitatively however, there is an important difference compared to the model at 800~K. Whereas at 800~K the photodissociation J$_1$ is responsible by more than 90$\%$ of the destruction of CO$_2$ is the upper atmosphere, at 1500~K, the contribution of the photodissociation is less important. At 0.1 mbar, $\sim$32$\%$ of the loss rate of CO$_2$ is due to the photodissociation J$_1$. Thus, even if this contribution is much more important than when one uses $\sigma_{CO_2}(\lambda, 300)$ (where it represents only 7$\times 10^{-4}\%$), one does not expect the increase of the loss rate of J$_1$ to be sufficient to impact significantly the abundance of CO$_2$ in this very hot atmosphere.

\begin{figure}[!htb]
\centering
\includegraphics[angle=0,width=\columnwidth]{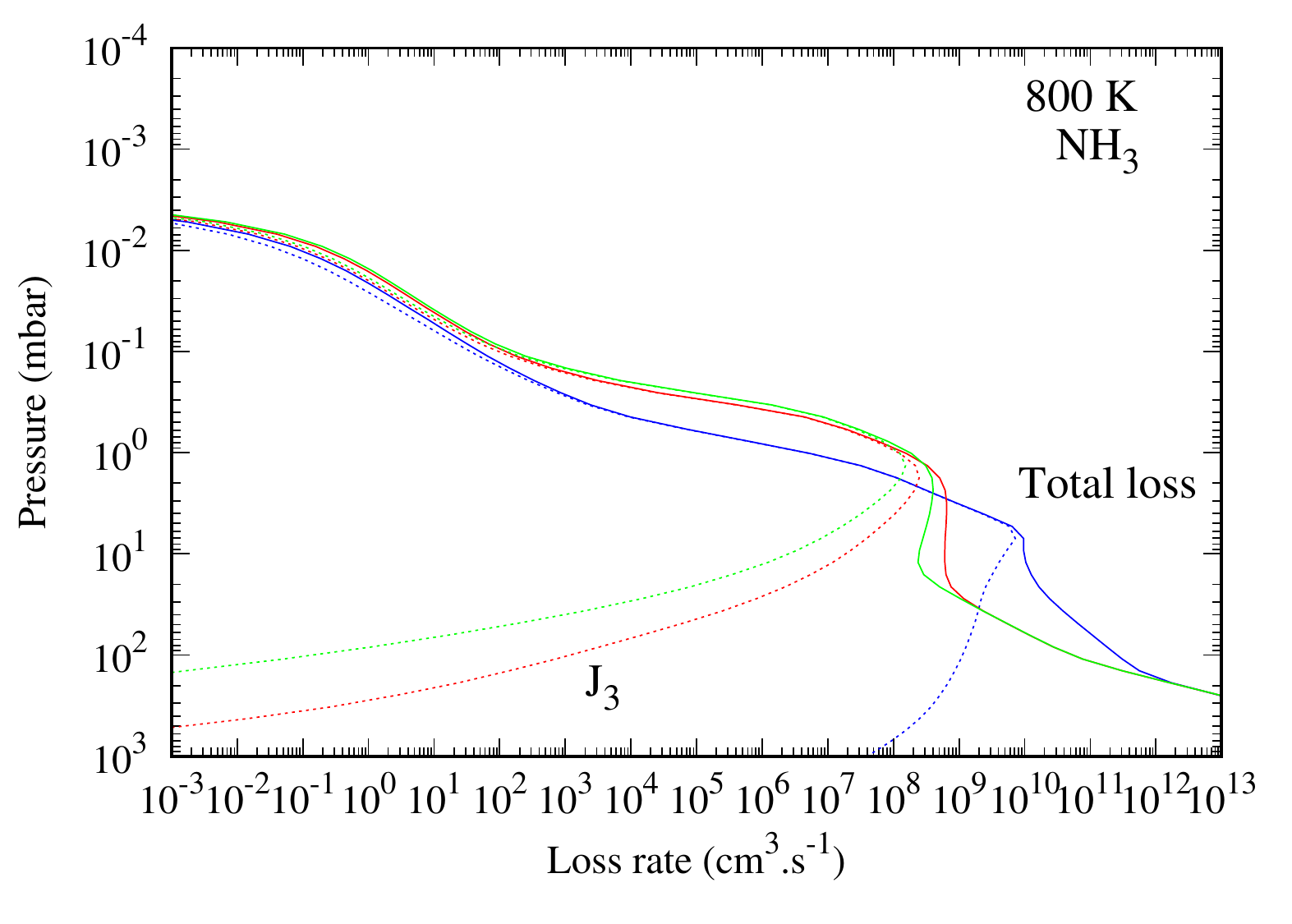}
\includegraphics[angle=0,width=\columnwidth]{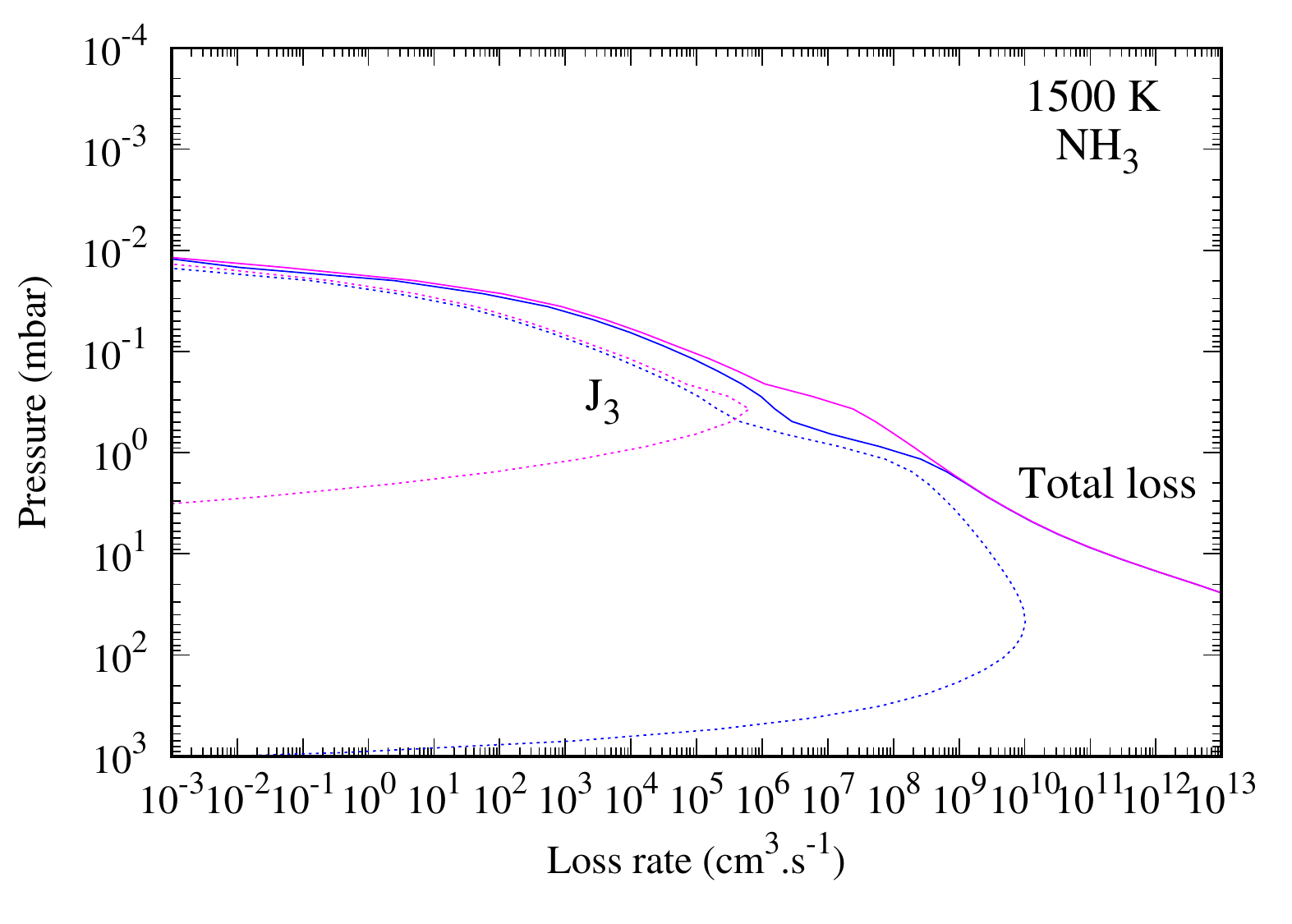}
\caption{For NH$_3$, total loss rates (\textit{full line}) and loss rates due to photolysis J$_3$ (\textit{dotted line}) in the two atmospheric models: "800~K" (\textit{top}) and "1500~K" (\textit{bottom}). Colors correspond to the different absorption cross section used: $\sigma_{CO_2}(\lambda, 300)$ (\textit{blue}), $\sigma_{CO_2}(\lambda, 800)$ (\textit{red}), $\sigma_{cont}(\lambda, 800)$ (\textit{green}), and $\sigma_{cont}(\lambda, 1500)$ (\textit{pink}).}
\label{fig:photo_rates_NH3}
\end{figure}

We see that using the warm absorption cross sections of CO$_2$ increases the role of photolysis in the destruction of carbon dioxide, which can even become, in some cases, the dominant processes in the high atmosphere. Consequently, the actinic flux is impacted (see Fig.~\ref{fig:pene}), which in return modifies the photodissociation rates of other absorbing molecules such as NH$_3$ and CH$_4$ (through Eq.~\ref{eq:J}, applied to other species). To illustrate this phenomenon of shielding, we present on Fig.~\ref{fig:photo_rates_NH3} the total loss rate of ammonia and the loss rate due to its single photodissociation route, NH$_3$ + $h\nu$ $\longrightarrow$ NH$_2$ + H (J$_3$). Although the absorption cross section of ammonia has not been modified between the different models, but only that of CO$_2$, the loss rates of NH$_3$ are impacted. With the hot absorption cross sections ($\sigma_{CO_2}(\lambda, 800)$, $\sigma_{cont}(\lambda, 800)$ and $\sigma_{CO_2}(\lambda, 1500)$), photodissociation of NH$_3$ occurs less deeper than with $\sigma_{CO_2}(\lambda, 300)$. For the atmosphere at 800~K, the maximum of J$_3$ loss rates is shifted from 7 mbar to $\sim$1.5 mbar, and from 50 mbar to 0.4 mbar for the atmosphere at 1500~K. Consequently, we expect that many molecules in addition of CO$_2$ will have their abundances modified. This phenomenon of shielding is further increased by the fact that the abundances of species are determined by a system of coupled differential equations. Thereby, each variation of abundance of one single species affects the species with which it reacts chemically.

\subsubsection{Chemical composition}

\begin{figure}[!ht]
\centering
\includegraphics[angle=0,width=\columnwidth]{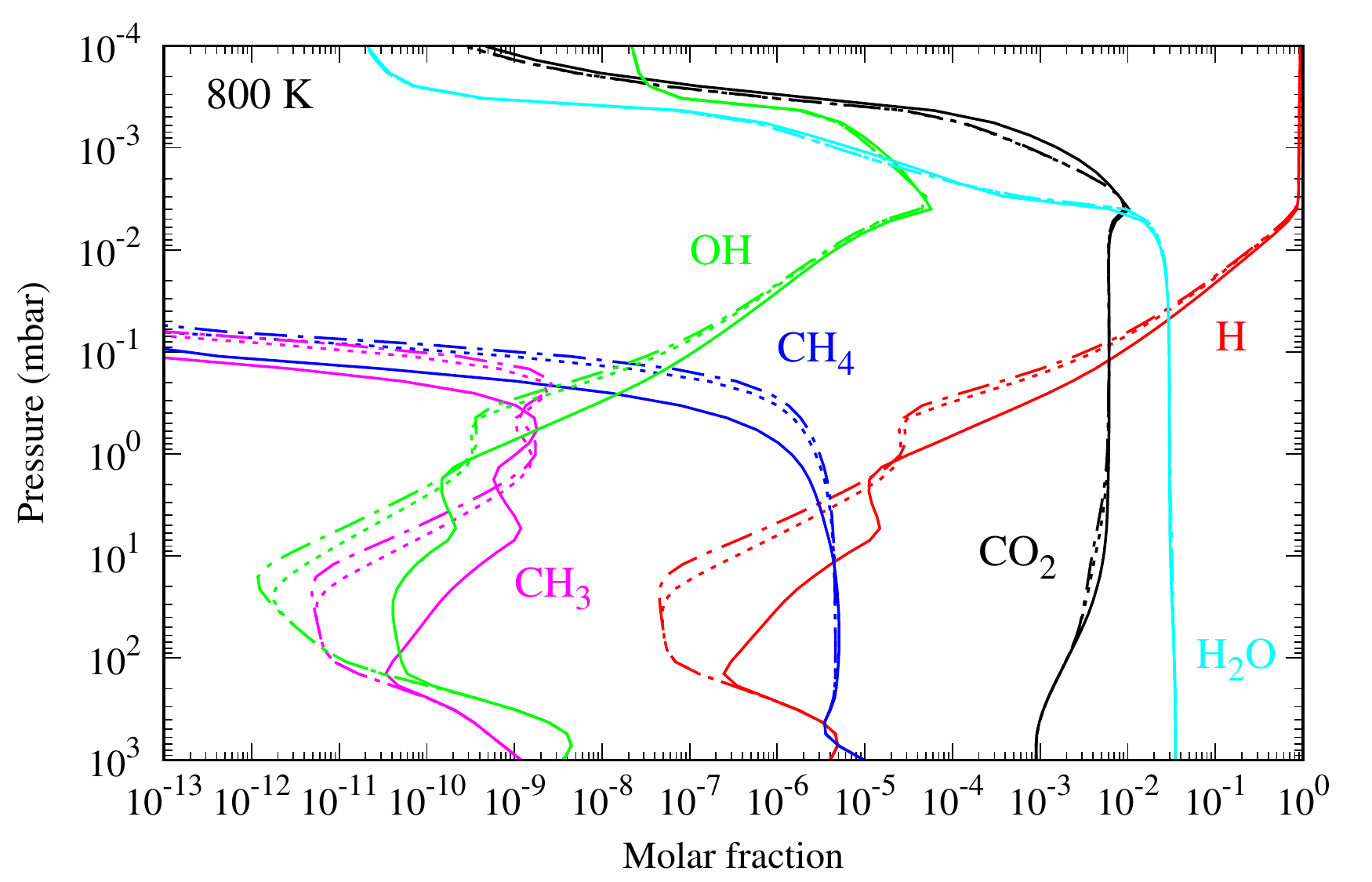}
\includegraphics[angle=0,width=\columnwidth]{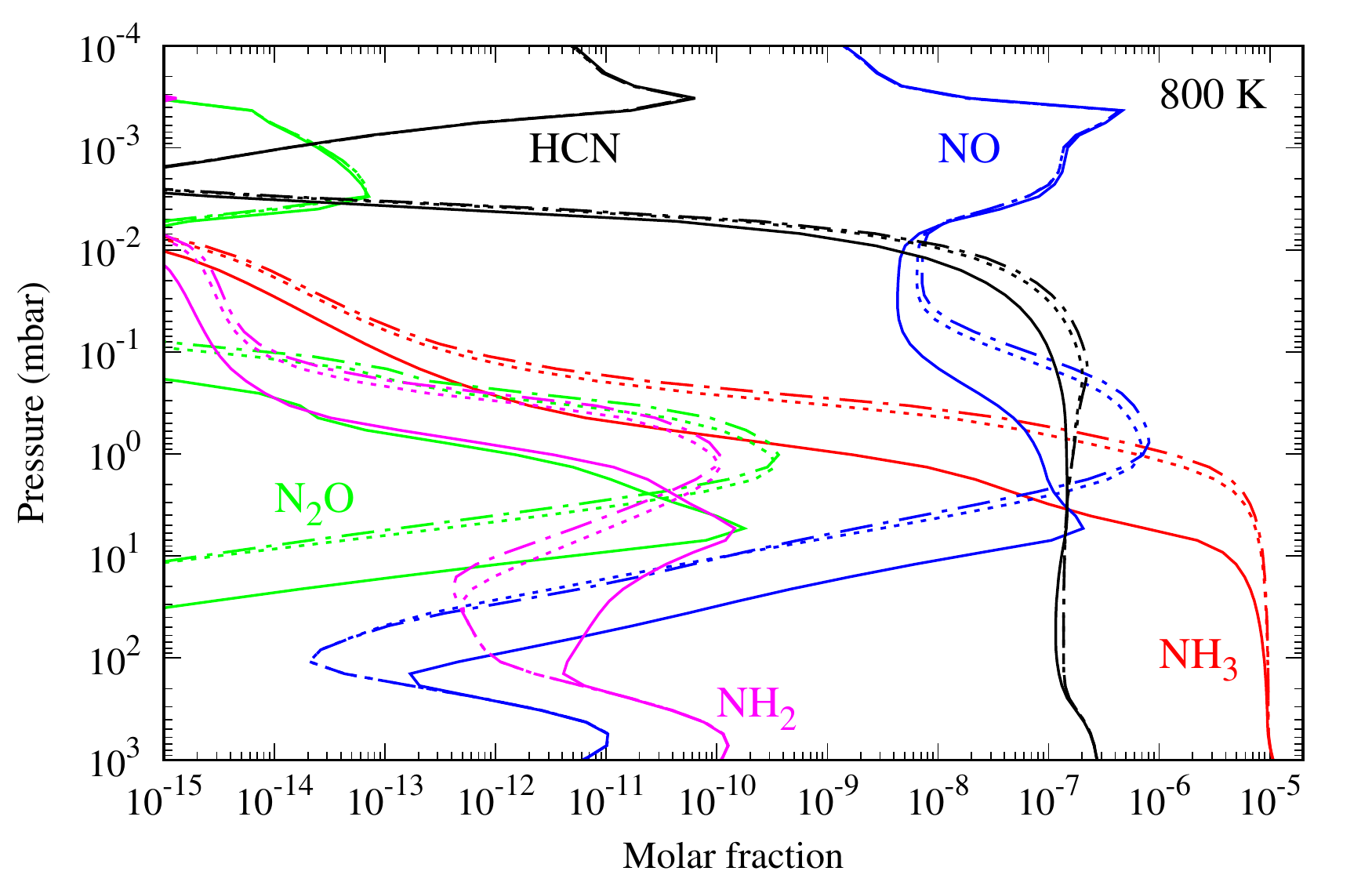}
\caption{Vertical mixing ratios of CO$_2$ and ten selected molecules computed with the photochemical model using $\sigma_{CO_2}(\lambda, 300)$ (full line), $\sigma_{CO_2}(\lambda, 800)$ (dotted line) and $\sigma_{cont}(\lambda, 800)$ (dotted-dashed line), for the thermal profile labeled~"800~K".}
\label{fig:abundances}
\end{figure}

For the 800 K atmosphere, we represented in Fig.~\ref{fig:abundances} the vertical mixing ratios of CO$_2$ and H$_2$O (as these species are the ones that contribute the more to the transmission spectra, see Sect.~\ref{sect:spectra}), as well as the mixing ratios of the nine molecules that are the most affected by the change of $\sigma_{CO_2}(\lambda, T)$, amongst species with a mixing ratio higher than $10^{-10}$. 
These results confirm what was expected looking at the loss rates: changing the absorption cross section of CO$_2$ affects the abundance of numerous species, not only carbon dioxide. The modification of the absorption cross section of carbon dioxide has an impact on the atmosphere down to pressure levels of 0.6~bar. Qualitatively, with $\sigma_{CO_2}(800, T)$  the vertical profiles are shifted to lower pressures. For a given pressure level, it results in lower abundances than with $\sigma_{CO_2}(300, T)$ for CO$_2$, H, and OH, and higher abundances for CH$_4$, NH$_3$, and HCN. The other species (CH$_3$, NO, NH$_2$, and N$_2$O) having tortuous vertical profiles, the direction of the change (i.e. increase or decrease) of the abundances depend on the pressure level. For NH$_3$, one can see the correspondance between the loss rate (Fig.~\ref{fig:photo_rates_NH3}) and the vertical profiles (Fig.~\ref{fig:abundances}). As it has been said in Sect.~\ref{sec:rates}, the maximum of J$_3$ loss rate varied from 7 mbar, with $\sigma_{CO_2}(300, T)$, to $\sim$1.5 mbar, with $\sigma_{CO_2}(800, T)$. These pressures correspond to the respective levels where the removal of ammonia by photolysis is not dominant anymore (if we go from low to high pressures) and NH$_3$ reaches an abundance of $\sim3\times10^{-6}$, close to its value at deeper levels.

\begin{figure}[!htb]
\centering
\includegraphics[angle=0,width=\columnwidth]{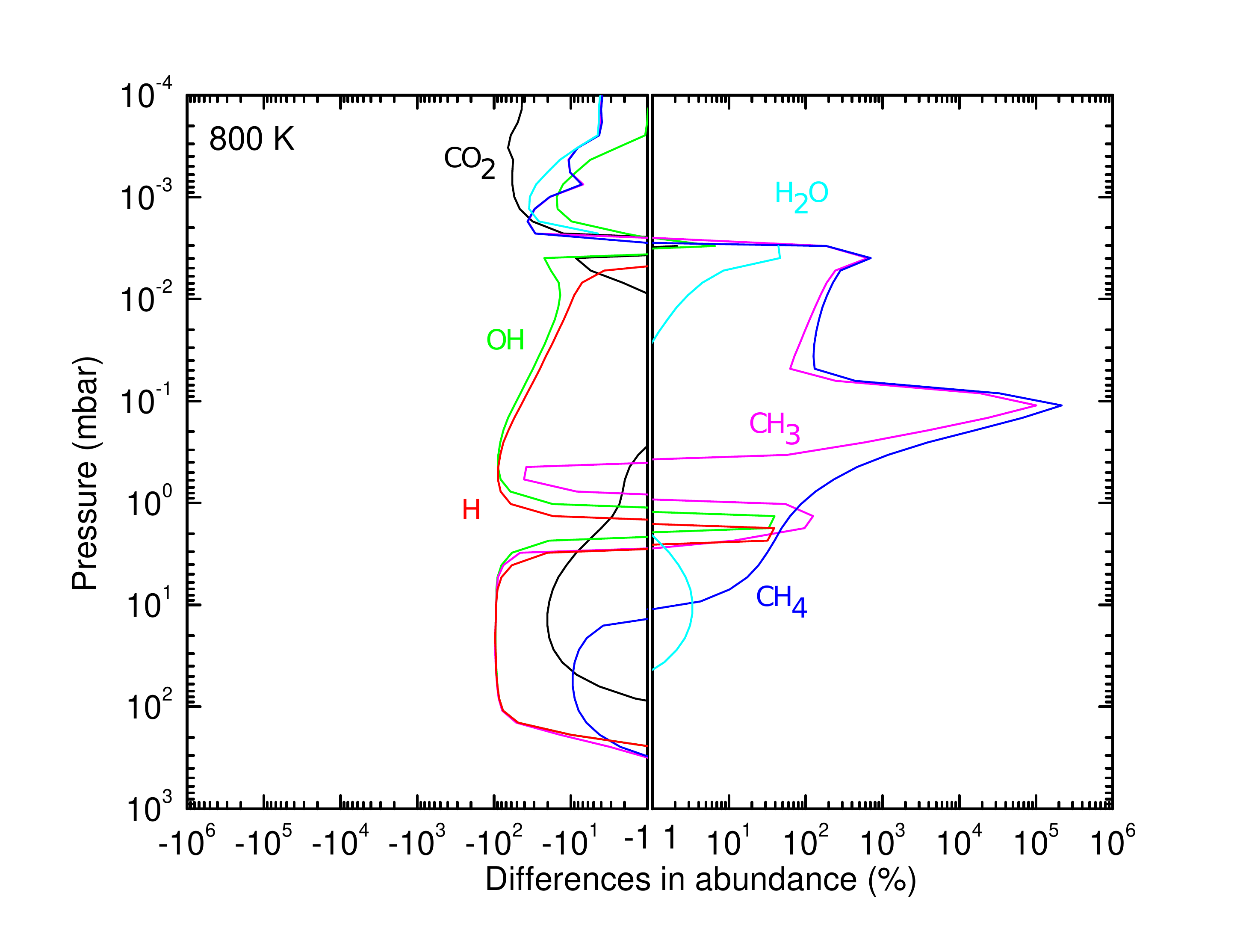}
\includegraphics[angle=0,width=\columnwidth]{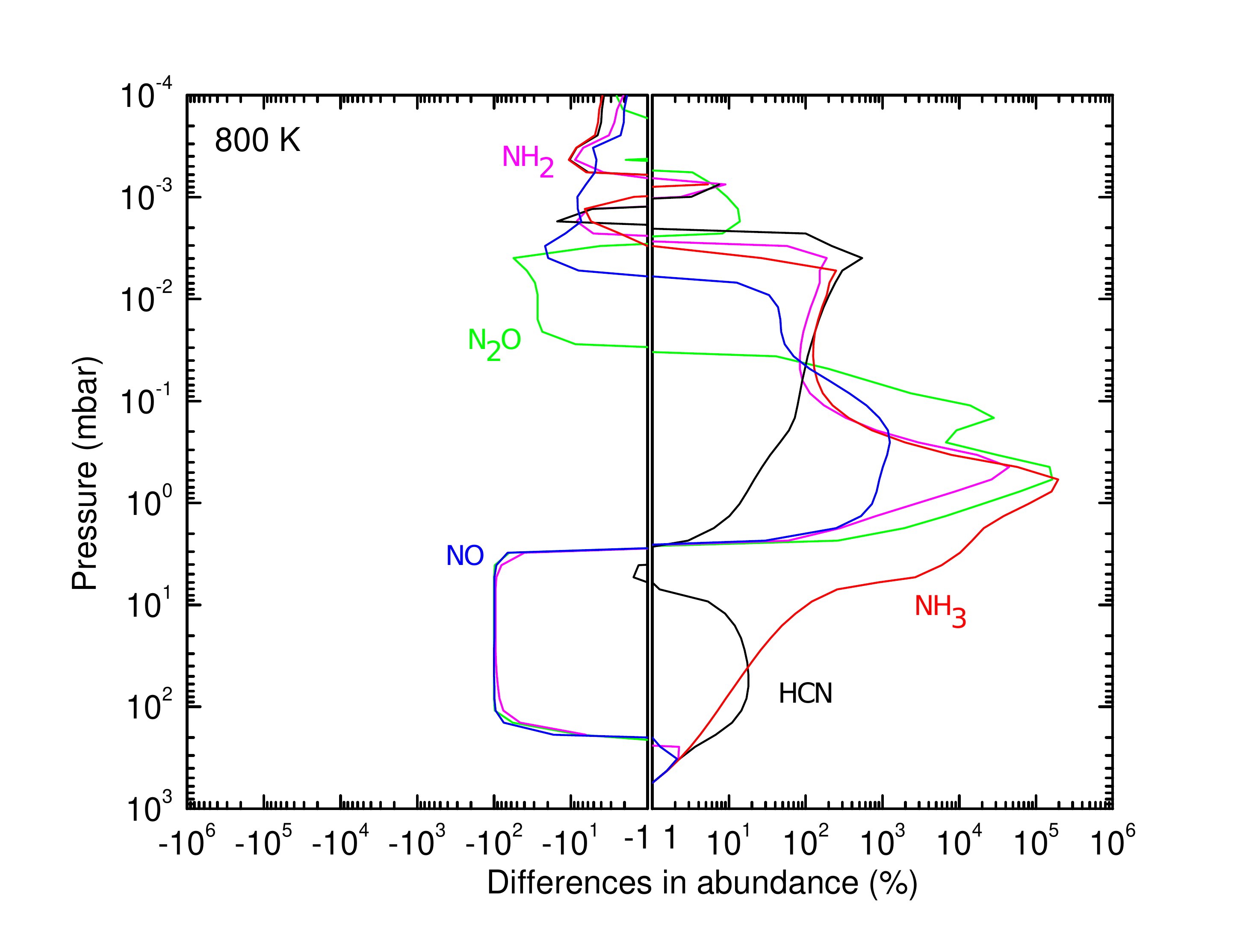}
\caption{Differences in abundances (\%) between the results obtained using $\sigma_{CO_2}(\lambda, 800)$ and $\sigma_{CO_2}(\lambda, 300)$, for the thermal profile labeled~"800~K".}
\label{fig:pourcentage}
\end{figure}

To quantify the change of composition induced by the warm CO$_2$ absorption cross section, we represented on Fig.~\ref{fig:pourcentage} the relative deviation of abundances (i.e. $((y^{800}_i-y^{300}_i)/y^{300}_i)\times100$, with $y^{T}_i$ the molar fraction of the species $i$ in the model using $\sigma_{CO_2}(\lambda, T)$) of the compounds shown in Fig.~\ref{fig:abundances} regarding the pressure level.
CH$_4$ and CH$_3$ show a large relative deviation of abundances at 0.1 mbar: $2.1\times10^5\%$ and $1\times10^5\%$ respectively. The abundance of NH$_3$, N$_2$O, and NH$_2$ are also strongly modified with relative deviation of $1.9\times10^5\%$, $1.6\times10^5\%$, and $3\times10^4\%$ respectively around 0.5 mbar.
These values are $\sim$150-300 $\times$ higher than the maximum relative deviation of CO$_2$ at $3.2\times 10^{-4}$ mbar (i.e. $-66\%$). Water is less affected with a maximum relative deviation of $-46\%$ at $4\times 10^{-3}$ mbar.\\
For the atmosphere at 800 K, we studied the deviation of the atmospheric composition induced by the use of the analytical formula (Eq.~\ref{eq:sigmatot}) applied with $T=800$ K, $\sigma_{cont}(\lambda, 800)$, instead of the experimental data $\sigma_{CO_2}(\lambda, 800)$. The atmospheric composition found with $\sigma_{cont}(\lambda, 800)$ is shown in Fig.~\ref{fig:abundances}. Deviations with the results obtained with $\sigma_{CO_2}(\lambda, 800)$ are negligible. We calculated the relative deviation of abundances (i.e. $((y^{800, cont}_i-y^{800}_i)/y^{800}_i)\times100$ and found a maximum for CH$_4$ (500$\%$ at 0.08 mbar). However, in the case studied here, $y_{CH4}$ is less than $10^{-11}$ at this pressure level. Thus, this deviation will have no impact on the observable corresponding to this composition, the planetary synthetic spectrum (cf. Sect.~\ref{sect:spectra}). \\

\begin{figure}[!ht]
\centering
\includegraphics[angle=0,width=\columnwidth]{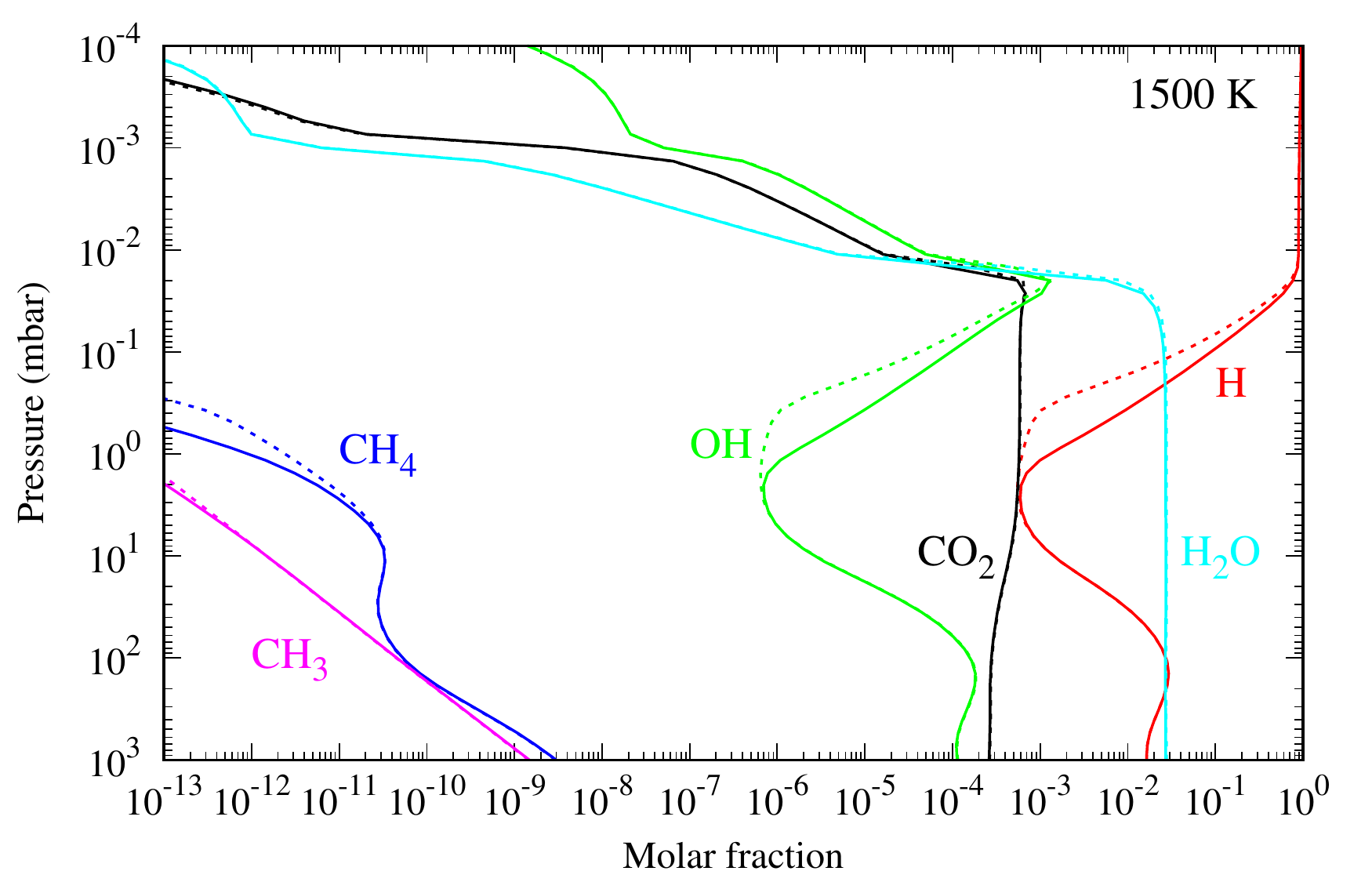}
\includegraphics[angle=0,width=\columnwidth]{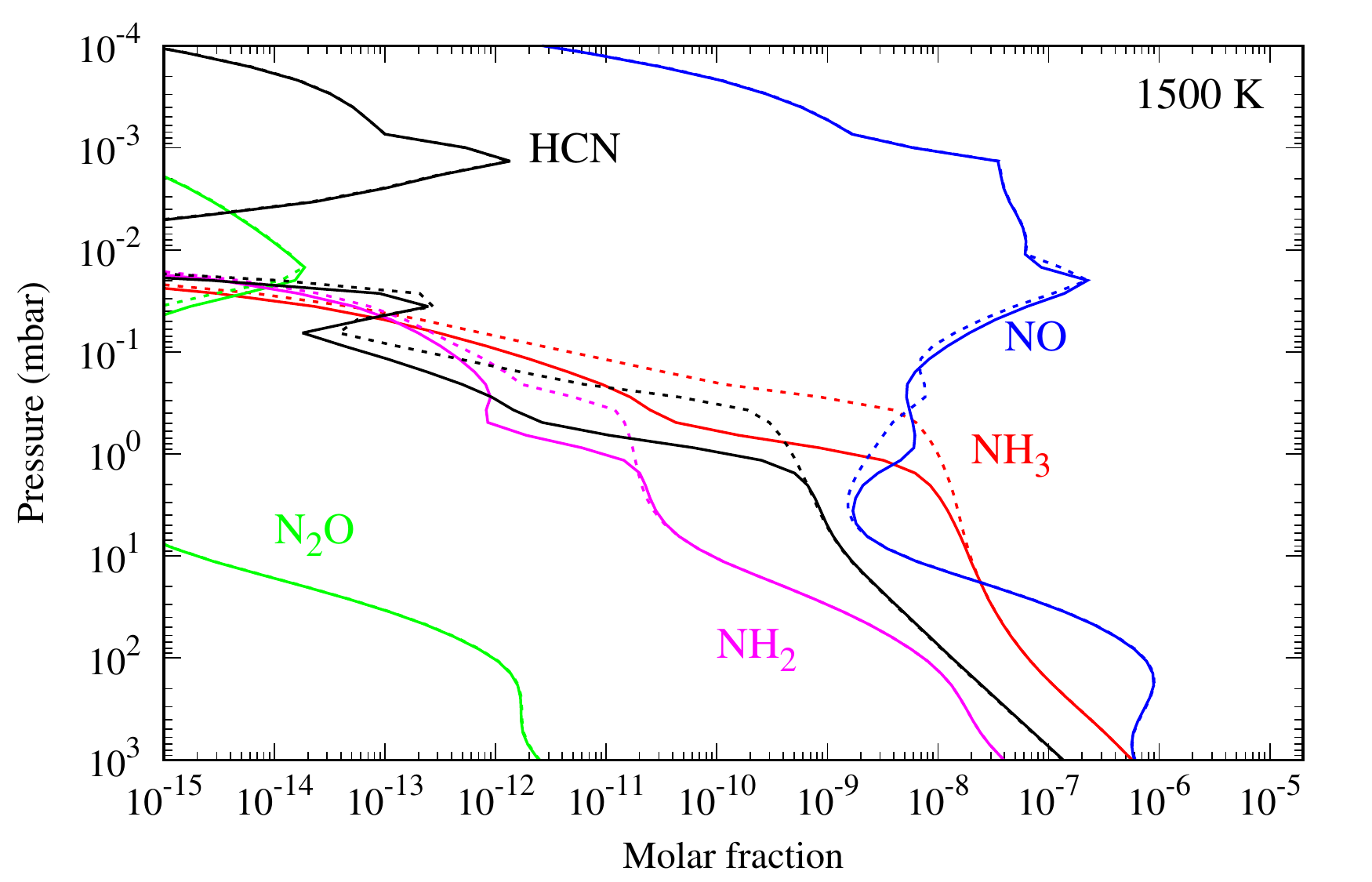}
\caption{Vertical mixing ratios of CO$_2$ and ten selected molecules computed with the photochemical model using $\sigma_{CO_2}(\lambda, 300)$ (full line) and $\sigma_{cont}(\lambda, 1500)$ (dotted line), for the thermal profile labeled~"1500~K".}
\label{fig:abundances_1500K}
\end{figure}

\begin{figure}[!htb]
\centering
\includegraphics[angle=0,width=\columnwidth]{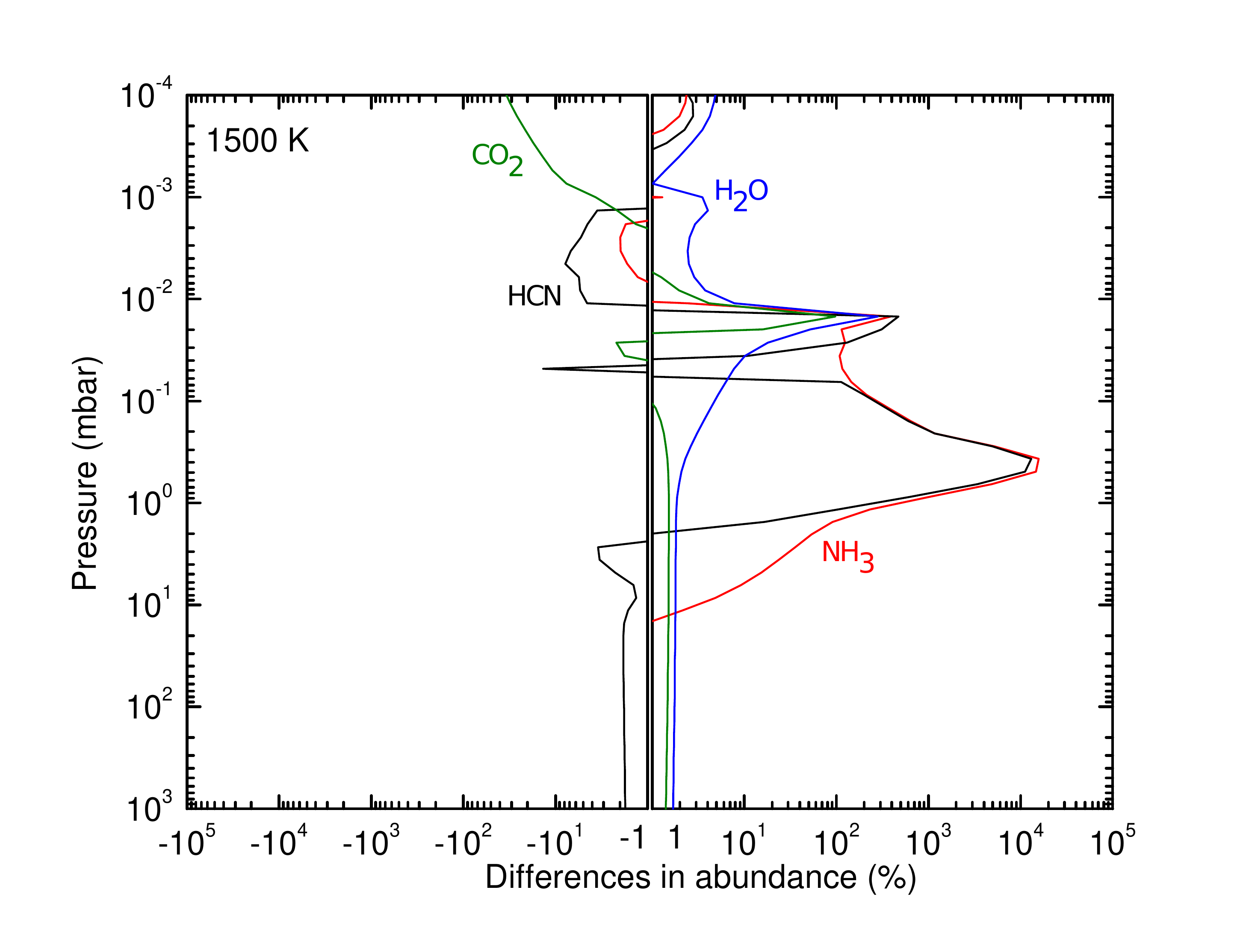}
\caption{Differences in abundances (\%) between the results obtained using $\sigma_{cont}(\lambda, 1500)$ and $\sigma_{CO_2}(\lambda, 300)$, for the thermal profile labeled~"1500~K".}
\label{fig:pourcentage1500}
\end{figure}

For the atmosphere at 1500~K, the use of $\sigma_{cont}(\lambda, 1500)$ instead of $\sigma_{CO_2}(\lambda, 300)$ modifies in a smaller extent the chemical composition (Fig.~\ref{fig:abundances_1500K}). The vertical abundance profile of CO$_2$ is barely modified.  Between 0.02 and 2 mbar, H and OH have lower abundances with the hot absorption cross section. As for the 800~K atmosphere, CH$_4$, NH$_3$, and HCN are destroyed less deeper by photolysis than with $\sigma_{CO_2}(\lambda, 300)$. Compared to the 800~K atmosphere, the abundance of NO is much less modified by the change of CO$_2$ absorption cross section.
Relative deviation of abundances (i.e. $((y^{1500, cont}_i-y^{300}_i)/y^{300}_i)\times100$ are represented in Fig.~\ref{fig:pourcentage1500}. Amongst species with mixing ratios $>$ $10^{-10}$, NH$_3$ and HCN are the ones that experience the largest variations. The relative deviation of their abundance is of 10$^4\%$ at 0.37 mbar. For, H$_2$O the relative deviation is of 281$\%$ at 1.4$\times 10^{-2}$ mbar. The other species present variations of less than 100$\%$.  

\subsubsection{Effect on observable}\label{sect:spectra}

\begin{figure}[!htb]
\centering
\includegraphics[angle=0,width=\columnwidth]{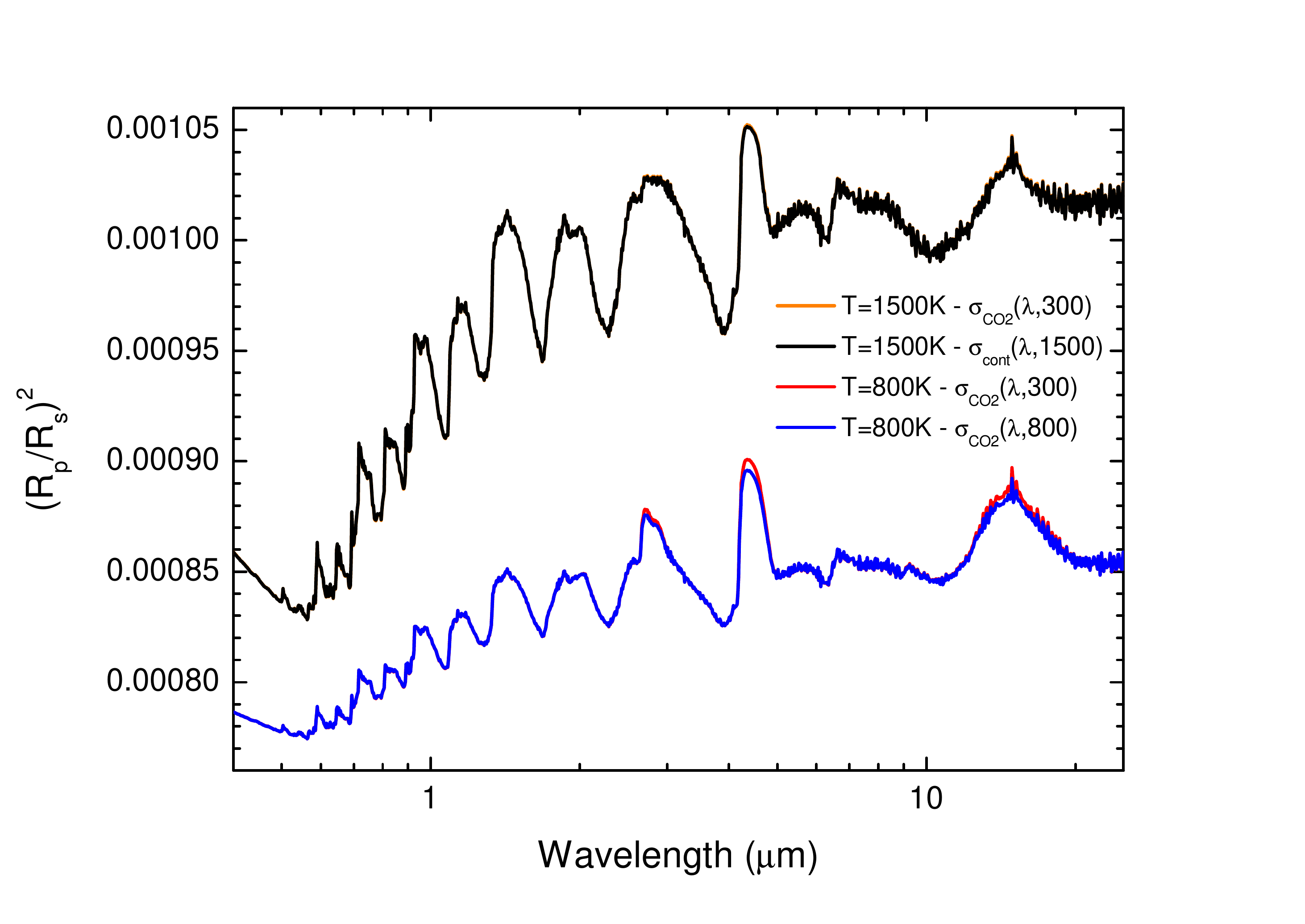}
\caption{Synthetic transmission spectra of the two warm Neptunes, corresponding to the chemical compositions calculated with $\sigma_{CO_2}(\lambda, 300)$ (\textit{red}) and $\sigma_{CO_2}(\lambda, 800)$ (\textit{blue}) for the "800~K" atmosphere and with $\sigma_{CO_2}(\lambda, 300)$ (\textit{orange}) and $\sigma_{cont}(\lambda, 1500)$ (\textit{black}) for the "1500~K" atmosphere. $R_p$ and $R_s$ are the planet and star radii, respectively. The spectrum shown was binned to a resolution (constant in $\lambda$) of R = 300.}
\label{fig:spectra}
\end{figure}

Infrared transmission spectra of the two warm Neptunes are presented in Fig.~\ref{fig:spectra}. For all of them, the absorption is dominated by H$_2$O except around 2.7, 4.3, and 14.9 $\mu$m, where CO$_2$ is the dominant absorbing species. We observed that the spectra of the "1500~K" atmosphere present larger features and are shifted to larger planetary radii, compared to the spectra of the colder atmosphere. This is due to the fact that the atmospheric scale height ($H=k_BT/\mu g$, where $\mu$ is the mean molecular weight) increases when the temperature increases, leading to a more extended atmosphere.

The differences between the spectra corresponding to the models using $\sigma_{CO_2}(\lambda, 300)$ and $\sigma_{CO_2}(\lambda, 800)$ are quite small. Variations are visible in the three CO$_2$ absorption bands. At 4.3 $\mu$m, the atmospheric absorption decreases by 5 ppm when using the warmer absorption cross section, which represents a decrease of 0.55$\%$. At 14.9 $\mu$m, a decrease of 4.8 ppm (0.55$\%$) is observed. Finally, a lower variation of 2.5 ppm (0.3 $\%$) is visible at 2.7 $\mu$m. 
These changes are below the level of uncertainty reached by recent observations of small planets in the size range of warm Neptunes/super-Earths (i.e. 60 ppm for \citealt{Kreidberg2014} and 22 ppm for \citealt{tsiaras2016b}). Reaching such a high level of sensitivity seems to be a real challenge even for the futur powerful JWST. Indeed, such a low S/N will be accessible but observations in this wavelength range will be in fact limited by systematics of the JWST, which are difficult to assess. They will probably be lower than the Hubble Space Telescope, i.e. about 20 ppm for the instruments NIRISS and NIRCAM \citep{rocchetto2016} and about 30 ppm for MIRI \citep{Beichman2014}.

However, one has to keep in mind that these deviations are only due to the change of the VUV absorption cross section of one single species. Using hot VUV absorption cross sections for all the species present in the atmosphere will probably have more consequences on the atmospheric composition and thus could produce larger effect on the transmission spectra.
\begin{figure*}[!htb]
\centering
\includegraphics[angle=0,width=\columnwidth]{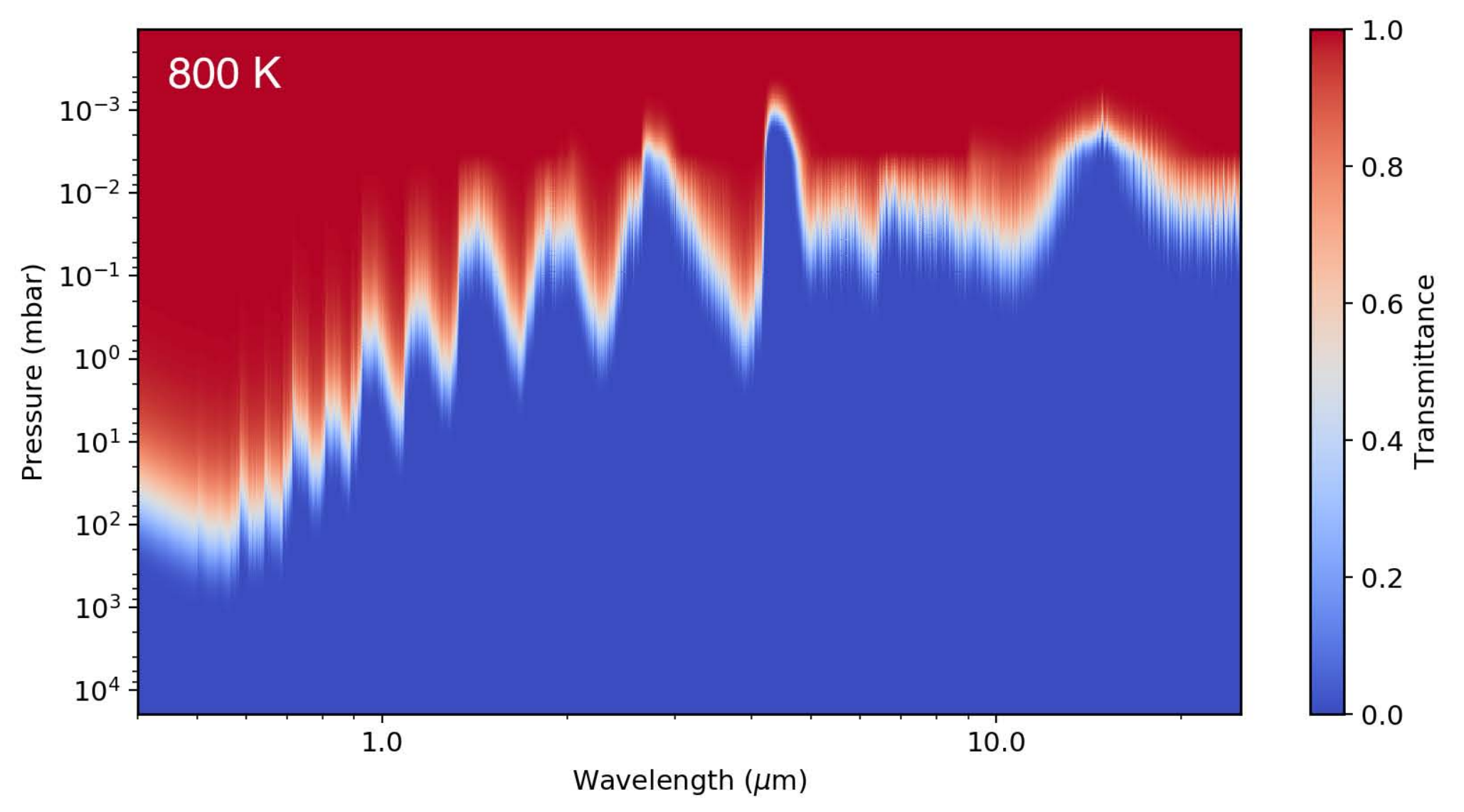}
\includegraphics[angle=0,width=\columnwidth]{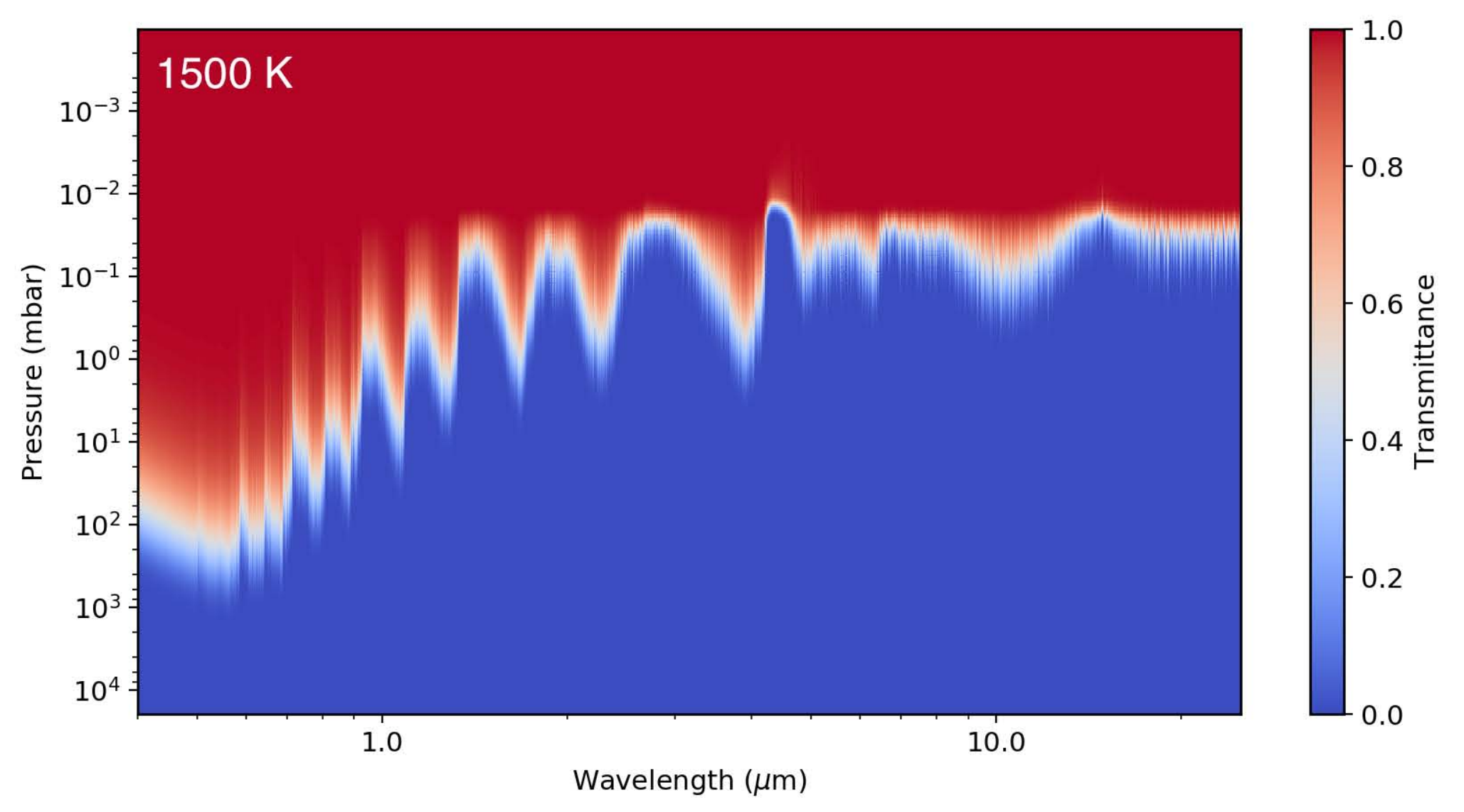}
\caption{Spectral transmittance as a function of pressure for the "800 K" (left) and the "1500 K" atmospheres (right). The plots correspond to the case where CO$_2$ absorption cross sections at respectively 800 K and 1500 K have been used.}
\label{fig:trans_level}
\end{figure*}
We also compared the spectra corresponding to the atmospheric compositions determined using $\sigma_{CO_2}(\lambda, 800)$ and $\sigma_{cont}(\lambda, 800)$. The variations are tiny. The maximum deviation occurs at 4.7 $\mu$m and is of 0.15 ppm (0.017$\%$). Such a level of variation confirms that using the analytical formula of the continuum of absorption instead of the real absorption cross section is a good approximation.

For the warmer atmosphere, the change of atmospheric composition due to the different VUV absorption cross sections of CO$_2$ almost does not affect the transmission spectrum. The most important deviation occurs at 4.7 $\mu$m and is of 1 ppm, which represents a decrease of 0.1 $\%$. This variation is not detectable with current instruments and it is highly probable that such a sensitivity will remain unreachable by the future ones during the next decades.

Figure \ref{fig:trans_level} represents the spectral transmittance as a function of pressure for the two atmospheres, in the case of the warm CO$_2$ absorption cross sections have been used. It allows us to see what are the pressure regions (and therefore the temperatures in Fig.~\ref{fig:PTprofile}) that are probed at different wavelengths for the two atmospheres. Generally, we notice that the same wavelength probes higher pressures in the 800 K atmosphere than in the 1500 K atmosphere. For the colder atmosphere, we see that the wavelengths for which we observe deviations between the spectra using $\sigma_{CO_2}(\lambda, 300)$ and $\sigma_{CO_2}(\lambda, 800)$, i.e. 2.5, 4.3 and 14.9 $\mu$m, correspond to the atmospheric layers between 10$^{-3}$ and 4$\times$10$^{-3}$ mbar, which corresponds to the pressure levels where carbon dioxide exhibits variation of $\sim$50\% and is more abundant than H$_2$O by more than one order of magnitude (Figs. \ref{fig:abundances} and \ref{fig:pourcentage}).
For the warmer atmosphere, the maximum variation between the spectra using different absorption cross sections is found at 4.7~$\mu$m. This wavelength probes the atmosphere around 10$^{-2}$ mbar, which corresponds to a level where the abundance of CO$_2$ varies by $\sim$100\%. Despite the larger magnitude of abundance deviation with respect to the 800 K atmosphere, the effect on the synthetic spectra is smaller. This is because CO$_2$ is less abundant (by a factor 10) in the warmer atmosphere and is about the same abundance than H$_2$O around 10$^{-2}$ mbar. Therefore the relative contribution of CO$_2$ to the planetary spectrum is lower than for the 800 K atmosphere.

\section{Conclusion}\label{sec:concl}

We present ten experimental measurements of the VUV absorption cross section of CO$_2$ from 150~K to 800~K on the wavelength range (115--230 nm), which allow us to quantify the temperature dependency of this data. We study more specifically the evolution of the continuum of absorption and determine a parameterization that depends only on the temperature and the wavelength. At temperatures higher than 500~K, using this parameterization is a good approximation for the absorption of CO$_2$, because the contribution of the fine structure on the absorption is less than 20~$\%$.

We study the implication of these new data for exoplanets studies using our thermo-photochemical model. In atmospheres around 800~K, using the appropriate absorption cross sections of CO$_2$ modifies the abundances of many species by several orders of magnitude. In our model, molecules that undergo most changes are CH$_4$, NH$_3$, N$_2$O, and CH$_3$. These changes lead to moderate deviations in transmission spectra (5 ppm) that would be hard to observe even with future instruments such as the James Webb Space Telescope. In warmer atmospheres ($\sim$1500~K), using of the appropriate CO$_2$ absorption cross sections has a lower impact on the atmospheric composition because the fast thermal kinetics dominates over the photodissociation processes in the area probed by observations. Thus, synthetic spectra are not impacted. We can therefore conclude that measurements at higher temperatures, larger than 1000~K, are not necessary in the context of warm exoplanet studies.

We also compare the results obtained using the real absorption cross section and the continuum calculated using the parameterization. Small differences are observed for the atmospheric composition but do not have visible implication for the synthetic spectra. This approximation can thus be used without any risk for atmospheres with temperatures higher than 500~K. For lower temperatures, the fine structure cannot be neglected, and a more detailed study on its variation with the temperature will be the subject of a forthcoming paper. 

It is worth noting that the changes observed in the atmospheric composition and the synthetic spectra are due to the change of only one parameter: the absorption cross section of carbon dioxide. Given the high coupling that exists between all the molecules through chemical kinetics and the phenomenon of shielding, it remains necessary to determine high temperature data for all absorbing species. Indeed, with the near launch of the James Webb Space Telescope (October 2018), reducing the uncertainty on the results of atmospheric models is paramount to be able to interpret correctly its future observations.

Data presented presented here are available in digital format through the CDS database.

\begin{acknowledgements} 
The authors wish to thank the anonymous referee for his very interesting comments. They also thank Gerd Reichard and Peter Baumg\"artel for their excellent assistance during the synchrotron radiation beam time periods. The authors also acknowledge the financial supports of the European Commission Programme "Access to Research Infrastructures" for providing access to the synchrotron facility BESSY in Berlin, of the programme PIR EPOV, and that of the CNRS/INSU Programme National de Plan\'etologie (PNP). O.V. acknowledges support from the KU Leuven IDO project IDO/10/2013, from the FWO Postdoctoral Fellowship programme, and from the Postdoctoral Fellowship programme of the Centre National d'Etudes Spatial (CNES). I.P.W. acknowledges support by the ERC project ExoLights (617119).
\end{acknowledgements}

\bibliographystyle{aa}
\bibliography{VENOT2017_CO2}

\end{document}